\documentclass[10pt,a4paper,twoside,bibliography=totocnumbered]{article}

\usepackage{xcolor}

\usepackage{graphicx}

\usepackage{amsmath}
\usepackage{amssymb}

\usepackage{geometry}

\usepackage{pdflscape}

\begin{document}

% title page
\title{{\huge \textcolor{black}{Combinatorial Potential of Random Equations with Mixture Models: Modeling and Simulation}\vspace*{-0.4cm}\\ \begin{minipage}{5cm}\centering \textcolor{black}{---}\vspace*{-0.4cm}
\end{minipage}\\}{\Large \textcolor{black}{Original Research Article}}\vspace*{0.5cm}}

%\subtitle{Bachelor Informatik}

\author{Wolfgang Hoegele$^{1}$\vspace*{0.4cm}\\
%{\normalsize Applied Mathematics in Medical Physics}\vspace*{0.4cm}\\
{\normalsize $^{1}$ Munich University of Applied Sciences HM}\\{\normalsize Department of Computer Science and Mathematics}\\{\normalsize Lothstraße 64, 80335 München, Germany}\vspace*{0.4cm}\\
{\normalsize corresponding mail: \texttt{wolfgang.hoegele@hm.edu}}\vspace*{0.4cm}}

\date{\today}

\maketitle
\thispagestyle{empty}

\textit{This manuscript is published and can be cited by}\medskip

\textit{W. Hoegele, Combinatorial potential of random equations with mixture models: Modeling and simulation,\\ Mathematics and Computers in Simulation, 239, pp. 696-715 (2026),\\ https://doi.org/10.1016/j.matcom.2025.07.033.}\medskip

\textit{This manuscript appears also on ArXiv.org\\
arXiv:2403.20152 [stat.CO],\\ https://doi.org/10.48550/arXiv.2403.20152}

\section*{Abstract}
The goal of this paper is to demonstrate the general modeling and practical simulation of random equations with mixture model parameter random variables. Random equations, understood as stationary (non--dynamical) equations with parameters as random variables, have a long history and a broad range of applications. The specific novelty of this explorative study lies on the demonstration of the combinatorial complexity of these equations and their solutions with mixture model parameters. In a Bayesian argumentation framework, we derive a likelihood function and posterior density of approximate solutions while avoiding significant restrictions about the type of nonlinearity of the equation or mixture models, and demonstrate their numerically efficient implementation for the applied researcher. In the results section, we are specifically focusing on expressive example simulations showcasing the combinatorial potential of random linear equation systems and nonlinear systems of random conic section equations. Introductory applications to portfolio optimization, stochastic control and random matrix theory are provided in order to show the wide applicability of the presented methodology.\medskip

\textbf{Keywords:} random equations, likelihood, nonlinear equations, portfolio optimization, stochastic control, random matrix theory\medskip

% table of contents
\newpage
\thispagestyle{empty}
\tableofcontents

\section*{About the Author}

Dr. Högele is a Professor of Applied Mathematics and Computational Science at the Department of Computer Science and Mathematics at the Munich University of Applied Sciences HM, Germany. His research interests are mathematical and stochastic modeling, simulation and analysis of complex systems in applied mathematics with specific applications to radiotherapy, imaging, optical metrology and signal processing.\medskip

\thispagestyle{empty}

\newpage

\section{Introduction}

In this explorative study, we are investigating (nonlinear) random equations with parameter random variables which are described by mixture model densities and demonstrate (i) how they can be modeled and (ii) how their probabilistic solution space can be efficiently calculated/simulated. The novel focus lies on the demonstration of the combinatorial potential contained in such random equations and we provide several expressive examples in order to illustrate this. In the following, we provide a broader context to this work.\smallskip

Finding solutions of nonlinear equation systems (if they exist) is in general a difficult task. Typical approaches are the reformulation for iterative methods, e.g. such as fixed--point equations with a contracting term, approximate numerical solutions based on Newton's method and many more \cite{Kelley1993}. This task is even further complicated if some parameters contain uncertainties described by random variables, since solutions (if they exist) will in general depend on the probability densities of these parameters. This leads to the field of Random Equations (RE), understood as stationary (non--dynamical) equations with parameters as random variables, which have a long history and a broad range of applications, such as in algebra (questions of roots of random algebraic equations) \cite{Kac1943,Wschebor2008}, econometrics \cite{Heckman1978,Masten2018}, parameter fitting of linear mixed--effects models in statistical software \cite{Bates2015} or inverse metrology problems such as phase retrieval utilizing quadratic equations \cite{Chen2016,Wang2018}. General results about the solution theory are going back to the 1960's and have been published for random (operator) equations \cite{Bharucha-Reid1964, Bharucha-Reid1972}. Although REs have a wide applicability, there is no general consolidated research field about REs but typically specific investigations for a restricted research field with major constraints about the type of equations or probability densities.\smallskip

A large body of current research in a related field covers stochastic dynamical systems by stochastic differential equations (SDEs) and their, not so popular but related, random differential equations (RDEs) \cite{Jornet_2023,Jornet_2023b,Jornet_2024}. In the latter, only parameters of the differential equation are described by random variables and no stochastic calculus (as for SDEs) is needed. Often RDEs are described by terms such as \textit{differential equations with uncertainties} \cite{Pulch_2024}, underlining the fact of missing a uniquely utilized label. We will follow the mindset of REs as the stationary (non--dynamical) analog of RDEs, i.e. the equations investigated contain parameter probability densities on which the solutions are depending on. {In this context, solving REs can also be interpreted as finding the distribution of stationary points for RDEs.}\smallskip

Mixture models are broadly applied in many research fields for describing given observations, which result from an overlay of several sub--populations. A typical task is to determine the distributions of the sub--populations with the widely applied \textit{Expectation Maximization} algorithms \cite[pp.519-544]{Gelman2013}. In the perspective of the author it is not a standard application to actively construct mixture models in order to describe general properties of a problem definition (not based on observations but problem inherent). In this study, we explore the consequences of such constructed mixture models in random equations. \smallskip

In order to explore the solution space of REs with mixture model parameters, we are focusing on approximate solutions, i.e. we will introduce a very weak form of solution based only on a likelihood argumentation. This allows straightforward formulations without significant restrictions about the type of nonlinear equation system and parameter densities.\smallskip

The style of this presentation is as follows: We present a general argumentation framework without requiring specific types of (nonlinear) equation systems, mixture model component types and their number, and dimensions. It is regarded as a main strength of this presentation that we focus on a broad applicability and an inclusive language for the applied researcher unfamiliar with REs. In the perspective of the author, a practical modeling approach for general REs along those lines is missing in the literature in order to understand the underlying complexity utilizing mixture model random variables. The closest presentations are by the author in the field of computer vision, utilizing mixture models in order to resolve object pose fitting problems for the case of unknown correspondences of object features \cite{Hoegele2024}, and in the field of semi-supervised error-in-variables regression in statistics, utilizing mixture models in order to work with only partially paired data \cite{Hoegele2025}.\smallskip

The Bayesian perspective in this study gets visible in two ways: a) We are actively constructing probability density functions, e.g. as mixture models, and do not consider them as a result of a stochastic process or observations, b) We are applying Bayes' theorem to the constructed likelihood function with a (possibly non-- or weakly informative) prior density in order to interpret the visual simulation result as posterior densities.\smallskip

The main presentation in the methods section follows this argumentation: i) Definition of REs with the parameters as random variables in Subsection \ref{sec:GenNom}, ii) Presentation of the formal calculation of the likelihood function and posterior density of the approximate solutions of the RE system in Subsection \ref{sec:Likelihood} (and \ref{sec:UsePrior}), iii) Investigating its combinatorial implications in Subsection \ref{sec:dirac}, iv) Introducing extensions to the general derivations in Subsection \ref{sec:extensions}, and v) Providing numerical remarks for an efficient implementation in Subsection \ref{sec:Numerics}.\smallskip

In the results section, we introduce a practical perspective on this type of modeling by providing expressive simulation examples for random linear equation systems (Subsection \ref{sec:res:randomlinear}), for nonlinear equations with random conic sections (Subsection \ref{sec:res:randomconic}), for a portfolio optimization analysis (Subsection \ref{sec:res:portfolio}), for a control engineering problem (Subsection \ref{sec:res:control}) as well as for an application in random matrix theory (Subsection \ref{sec:res:matrix}).

\section{Methods}

\subsection{General Nomenclature}
\label{sec:GenNom}

Probability density functions of random variables are denoted by $\boldsymbol{A} \sim f_{\boldsymbol{A}}$ or $\boldsymbol{B} \sim f_{\boldsymbol{B}}$. We assume bounded and Lebesgue integrable density functions $f_{\boldsymbol{A}}\in L^1(\mathbb{R}^K)$ or $f_{\boldsymbol{B}}\in L^1(\mathbb{R}^R)$, which contains broadly applied standard cases such as multivariate normal or uniform distributions.  Note, also Dirac distributions $\delta$ as a limit of such density functions in the line of technical argumentation are considered, but measure theoretic discussions are not focus of this work and it will be referred to literature for such a standard treatment \cite{Bharucha-Reid1972}.\smallskip

In the context of this work, we define a \textit{random equation} or \textit{random equation system} by
\begin{align}
\boldsymbol{M}(\boldsymbol{x};\boldsymbol{A}) &= \boldsymbol{B}  \;.\label{equ:genequsys}
\end{align}
with $\boldsymbol{M} : \mathbb{R}^n \times \mathbb{R}^K \rightarrow \mathbb{R}^R$ a nonlinear real vector-valued function (containing $R$ components $M_r:\mathbb{R}^n \times \mathbb{R}^K \rightarrow \mathbb{R}$, $r=1,..,R$), $\boldsymbol{A}\sim f_{\boldsymbol{A}}$ the random variable vector of the parameters of the left hand side with $K$ entries and $\boldsymbol{B}\sim f_{\boldsymbol{B}}$ the random variable vector on the right hand side with $R$ entries.  {In this work, we assume $\boldsymbol{A}$ and $\boldsymbol{B}$ to be \textit{independent} random variables.} The vector $\boldsymbol{x}\in\mathbb{R}^n$ is the unknown solution of the equation system. In this definition $\boldsymbol{x}$ is not explicitly introduced as a random variable, however, the probabilistic solution $\boldsymbol{x}$ depends on the parameter random variables and this, consequently, makes $\boldsymbol{x}$ a random variable in the Bayesian interpretation of this work. The choice of considering two separate random variable vectors $\boldsymbol{A}$ and $\boldsymbol{B}$ is due to its beneficial argumentation in Subsection \ref{sec:Likelihood} and extensions to avoid this are presented in Subsection \ref{sec:implicit}.\smallskip

The main technical restriction on $\boldsymbol{M}$ is that the composite function $f_{\boldsymbol{B}}(\boldsymbol{M}(\boldsymbol{x};\cdot))\, f_{\boldsymbol{A}}(\cdot)\in L^1(\mathbb{R}^K)$ remains Lebesgue integrable for all $\boldsymbol{x}\in\mathbb{R}^n$, in order to be able to apply the \textit{law of total probability} for  density functions in the general derivation in Subsection \ref{sec:Likelihood}. This is regarded as a very weak constraint for practical applications such as presented in the results Section \ref{sec:results} considering $f_{\boldsymbol{B}}(\boldsymbol{M}(\boldsymbol{x};\cdot))$ being bounded and $f_{\boldsymbol{A}}$ Lebesgue integrable.\smallskip

\textit{Nonlinearity} of $\boldsymbol{M}$ in $\boldsymbol{x}$ is understood as a general label, i.e. either it is linear (i.e. additivity and homogeneity with respect to $\boldsymbol{x}$ is fulfilled) or strictly nonlinear (at least one property is not fulfilled). This means, we are not excluding linear equations (as we will demonstrate insightful examples in the results Section \ref{sec:results} also for linear equation systems), but only mark that linearity is neither guaranteed nor necessary in the derivations of this work.\smallskip

We will utilize the phrases \textit{solution} or \textit{solution space} of a random equation only in an approximate sense, and not as specific (\textit{wide--} or \textit{strict--sense}) solutions which are a well studied subject \cite{Bharucha-Reid1972}. This means, we regard the problem completely in the likelihood perspective of estimation theory, i.e. which $\boldsymbol{x}$ fulfill the Equation (\ref{equ:genequsys}) most likely. This is a much weaker perspective on the solution space, since it allows solutions in a stricter sense \cite{Bharucha-Reid1972} but also approximate \textit{solutions} in one framework. The benefit is that this allows straightforward argumentations in Subsection \ref{sec:Likelihood}, on the other hand, we are not able to distinguish between approximate solutions and solutions in a stricter sense. In the perspective of an applied researcher, we regard these approximate solutions as a sufficient approach for many practical applications, and provide example investigations based on this understanding in the numerical results Section \ref{sec:results}.\smallskip

In the scope of this work, we mainly utilize random variables with \textit{mixture model densities composed of separated/disjoint components} in order to provide combinatorial insights when applied to random equations (but overlaps of mixture model components are explicitly not excluded). This can be interpreted as to find approximate solutions for many combinatorial different random equations simultaneously, i.e. working with multi--valued parameters. Such equations might occur in situations where parameters of the equations itself are estimated and/or measured with significant uncertainties or deliberately one wants to allow several possibilities simultaneously for parameters as part of the problem definition. In this respect, similarities of such modeling approaches can be observed in the construction of Dirac combs (and related expressions) in information theory in discretization strategies \cite[pp. 125-126]{Strichartz2003}.

\subsection{General Derivation of the Likelihood Function and Posterior Density}
\label{sec:Likelihood}
We reformulate Equation (\ref{equ:genequsys}) by focusing on the difference random variable $\boldsymbol{M}(\boldsymbol{x};\boldsymbol{A}) - \boldsymbol{B}$ for the definition of a likelihood function $\mathcal{L}$. We interpret this as follows: Find the approximate solutions $\boldsymbol{x}$ that makes the value $\boldsymbol{0}$ for the difference random variable most likely.
\begin{align*}
\mathcal{L}(\boldsymbol{0}\,\vert\,\boldsymbol{x}) \quad=\quad&  f_{ \boldsymbol{M}(\boldsymbol{x};\boldsymbol{A}) - \boldsymbol{B}}(\boldsymbol{0})\\
\text{\small (law of total probability)}\quad{=}\quad& \int\limits_{\mathbb{R}^{K}}  f_{\boldsymbol{M}(\boldsymbol{x};\boldsymbol{s}) - \boldsymbol{B}}(\boldsymbol{0})\cdot f_{\boldsymbol{A}}(\boldsymbol{s})\;\text{d}\boldsymbol{s}\\
\text{\small (shifted $\boldsymbol{B}$)}\quad{=}\quad& \int\limits_{\mathbb{R}^{K}} f_{\boldsymbol{B}}(\boldsymbol{M}(\boldsymbol{x};\boldsymbol{s}))\cdot f_{\boldsymbol{A}}(\boldsymbol{s})\;\text{d}\boldsymbol{s}
\end{align*}
This is a closed form representation of the likelihood function. Utilizing the law of total probability for probability densities, $f_{\boldsymbol{M}(\boldsymbol{x};\boldsymbol{s}) - \boldsymbol{B}}$ is a conditional density with the condition $\boldsymbol{A}=\boldsymbol{s}$. 

If we further assume that all entries {of the random variable vector $\boldsymbol{B}$, i.e. $B_r$ for $r=1,..,R$,} are stochastically independent, we get
\begin{align}
=\quad& \int\limits_{\mathbb{R}^{K}} \left(\prod\limits_{r=1}^R\;  f_{B_r}(M_r(\boldsymbol{x};\boldsymbol{s}))\right)\cdot f_{\boldsymbol{A}}(\boldsymbol{s})\;\text{d}\boldsymbol{s}\;.\label{equ:Likelihood_1}
\end{align}
Next, if we further assume that all entries of the random variable vector $\boldsymbol{A}${, i.e. $A_i$ for $i=1,..,K$,} are stochastically independent, we get
\begin{align}
\quad{=}\quad& \int\limits_{\mathbb{R}^{K}} \left(\prod\limits_{r=1}^R\;  f_{B_r}(M_r(\boldsymbol{x};\boldsymbol{s}))\right)\cdot \left(\prod\limits_{i=1}^{K}\;f_{A_{i}}(s_i)\right)\;\text{d}\boldsymbol{s}\;.\label{equ:Likelihood_2}
\end{align}
{At this point it is important to note, that this is the formula on which the practical calculation of the approximate solutions $\boldsymbol{x}$ will be based on, also for the case with mixture model densities (see for more on this in the Numerical Remarks in Subsection \ref{sec:Numerics}). The further derivations by inserting mixture models densities directly into this formula (especially in Subsection \ref{sec:dirac}) will demonstrate the combinatorial complexity that is contained in these equations, which helps in the theoretical interpretation of the results.}\medskip

By applying the standard definition of the posterior applying Bayes' theorem, we get
\begin{align*}
\pi(\boldsymbol{x}\,\vert\,\boldsymbol{0}) \propto \mathcal{L}(\boldsymbol{0}\,\vert\,\boldsymbol{x})\cdot \pi(\boldsymbol{x})\;,
\end{align*}
with the prior density function $\pi$ of the random variable $\boldsymbol{X}$. Note, essentially non-- or weakly informative priors (such as $\pi(\boldsymbol{x})=$\textit{const.} on a finite, but large or problem specific domain) will be utilized (more about this in Subsection \ref{sec:UsePrior}) in order to not influence the solution process by subjective information. In consequence, the posterior density function can be calculated to (by neglecting positive normalization constants)
\begin{align}
\pi(\boldsymbol{x}\,\vert\,\boldsymbol{0})\propto& \left(\;\int\limits_{\mathbb{R}^{K}} \left(\prod\limits_{r=1}^R\;  f_{B_r}(M_r(\boldsymbol{x};\boldsymbol{s}))\right)\cdot \left( \prod\limits_{i=1}^{K}\;f_{A_{i}}(s_i)\right)\;\text{d}\boldsymbol{s}\right)\cdot \pi(\boldsymbol{x})\label{equ:posterior_1}\;.
\end{align}
Main part of this simulation study is to investigate the solutions of Equation (\ref{equ:posterior_1}) for different types of equation systems and density functions of the random variables. {The importance of introducing the posterior density based on the likelihood function is as follows: The shape of the intensity map of the likelihood function does not tell directly about the contained uncertainties about the approximate solution. In order to achieve this, we need to transfer such likelihood intensity maps to true probability densities such as the posterior, since this allows us to calculate probabilities, for example, in the form of \textit{credibility regions}. Although this is only a standard step in Bayesian argumentation, it is essential in order to assess the quality of the solutions found directly in the intensity maps. Further, in the Results Subsections \ref{sec:res:randomlinear} and \ref{sec:res:portfolio} we demonstrate the application of nontrivial priors.}\medskip

By utilizing, mixture models for the parameter random variable vectors $\boldsymbol{A}$ and $\boldsymbol{B}$, we attempt to find a suitable $\boldsymbol{x}$ which solves the stochastically modeled equation systems with multiple modes in their densities. If we describe the mixture models by 
\begin{align*}
A_{i}\sim \sum\limits_{j=1}^{L_{A,i}} v_j\cdot f_{A_{i,j}},\quad B_{r}\sim \sum\limits_{j=1}^{L_{B,r}} w_j\cdot f_{B_{r,j}}
\end{align*}
with given mixture model component weights $\sum\limits_{j=1}^{L_{A,i}} v_j = 1$ and  $\sum\limits_{j=1}^{L_{B,r}} w_j = 1$ for $v_j,w_j>0$ (typically, we will use in this work $v_j=\frac{1}{L_{A,i}}$ and $w_j=\frac{1}{L_{B,r}}$ since we are mainly interested in combinatorial investigations) and mixture model component densities $f_{A_{i,j}}$ for each random variable $A_{i}$ and $f_{B_{r,j}}$ for each random variable $B_r$, the formula of the Likelihood results in
\begin{align}
\mathcal{L}(\boldsymbol{0}\,\vert\,\boldsymbol{x}) =\quad& \int\limits_{\mathbb{R}^{K}}  \left[ \prod\limits_{r=1}^R\; \left( \sum\limits_{j=1}^{L_{B,r}} w_j\cdot f_{B_{r,j}}(M_r(\boldsymbol{x};\boldsymbol{s}))\right)\right]\cdot \left[\prod\limits_{i=1}^{K}\; \left( \sum\limits_{j=1}^{L_{A,i}} v_j\cdot f_{A_{i,j}}(s_i)\right)\right]\;\text{d}\boldsymbol{s}\;.\label{equ:Likelihood_3}
\end{align}
In this compact written form with two products over sums lies the essential enabler of high combinatorial power investigated in this study, which will be further demonstrated in Subsection \ref{sec:dirac}. The relevance of this presentation will be that no calculation of all possible realizations of combinations of parameters of the equation systems is needed, since this is taken care of by working directly with the mixture model densities.\medskip

\textbf{Remark:} The deeper understanding of the occurrence of the high combinatorial power lies in the understanding that the expression on the left hand side of Equation (\ref{equ:explaincompow}) with $R\cdot (V-1)$ additions and $R-1$ multiplications leads after the expansion to 
\begin{align}
\prod\limits_{r=1}^R \left( \sum\limits_{v=1}^{V} \alpha_{r,v} \right) = \sum\limits_{\boldsymbol{q}\,\in Q} \prod\limits_{r=1}^R \alpha_{r,q_r}\label{equ:explaincompow}
\end{align}
with the sum over $\#{}Q:=V^R$ combinations of the index tuples $\boldsymbol{q}:=(v_1,..,v_R)\in Q$ and $v_r\in\{1,..,V\}$ with $V^R-1$ additions and $V^R\cdot (R-1)$ multiplications. For example, having moderate numbers, such as $V=6$ and $R=10$, this leads to $V^R=60466176$ combinations of the $\alpha_{r,v}$ considered in the compact left hand side expression. It is an essential observation that in Equation (\ref{equ:Likelihood_3}) such expressions appear twice.

\subsection{Combinatorial Investigation with Dirac Distributions}
\label{sec:dirac}

{In the following we want to theoretically investigate the combinatorial complexity contained in these equations when using mixture model random vectors.} Assuming there are no uncertainties in the equation system, we can identify this by $A_{i}\sim\delta_{a_{i}}$ and $B_r\sim\delta_{b_r}$ as Dirac distributions \cite{Bharucha-Reid1972}. Please note, in this context we will chose the weakly informative prior $\pi(\boldsymbol{x})=$\textit{const.} on a large enough domain. We get
\begin{align*}
\pi(\boldsymbol{x}\,\vert\,\boldsymbol{0})\propto& \prod\limits_{r=1}^R\; f_{B_r}(M_r(\boldsymbol{x};\boldsymbol{a}))
\end{align*}
by neglecting the constant prior and applying the \textit{sifting property} of the Dirac distribution. This directly implies that the only non--zero probability density is at
\begin{align*}
M_r(\boldsymbol{x};\boldsymbol{a}) = b_r\quad\forall r=1,..,R\;,
\end{align*}
which is exactly the deterministic version of the equation system. An interesting theoretical extension can be demonstrated if the random variables are composed of mixtures of Dirac distributions
\begin{align*}
A_{i}\sim \frac{1}{L_{A,i}}\sum\limits_{j=1}^{L_{A,i}} \delta_{a_{i,j}},\quad B_{r}\sim \frac{1}{L_{B,r}}\sum\limits_{j=1}^{L_{B,r}} \delta_{b_{r,j}}\;.
\end{align*}
Then we get for the posterior distribution in a first step (again neglecting the constant prior)
\begin{align*}
\pi(\boldsymbol{x}\,\vert\,\boldsymbol{0})\propto& \int\limits_{\mathbb{R}^{K}} \left( \prod\limits_{r=1}^R\;  f_{B_r}(M_r(\boldsymbol{x};\boldsymbol{s}))\right)\cdot \left(\prod\limits_{i=1}^{K}\;\left(\frac{1}{L_{A,i}}\sum\limits_{j=1}^{L_{A,i}} \delta_{a_{i,j}}(s_i) \right)\right)\;\text{d}\boldsymbol{s}\;.
\end{align*}
The product of the sum corresponding to $\boldsymbol{A}$ inside the integral takes care of all combinations $\#{}P:=\prod\limits_{i=1}^{K}L_{A,i}$ of these discrete options $\boldsymbol{a}_{\boldsymbol{p}}:=(a_{1,p_1},\dots,a_{K,p_K})\in P$ and $p_i \in \{1,\dots,L_{A,i}\}$ of the entries of $\boldsymbol{A}$. By expanding the product inside the integral to a large sum, applying the sifting property of the Dirac distribution to each integral and neglecting positive constants, we get
\begin{align*}
\pi(\boldsymbol{x}\,\vert\,\boldsymbol{0})\propto&  \sum_{\boldsymbol{p}\,\in P} \left( \prod\limits_{r=1}^R \,f_{B_r}(M_r(\boldsymbol{x};\boldsymbol{a}_{\boldsymbol{p}})) \right)\;,
\end{align*}
with $\boldsymbol{a}_{\boldsymbol{p}}$ representing one of the $\#{}P$ combinations of the discrete values $a_{i,j}$ in each entry of $\boldsymbol{a}$. In a second step, inserting the definition of $f_{B_r}$ we get
\begin{align*}
=&  \sum_{\boldsymbol{p}\,\in P} \left( \prod\limits_{r=1}^R\,\left(\frac{1}{L_{B,r}}\sum\limits_{j=1}^{L_{B,r}} \delta_{b_{r,j}}(M_r(\boldsymbol{x};\boldsymbol{a}_{\boldsymbol{p}}))\right) \right)\;,
\end{align*}
which can further be interpreted (by neglecting positive constants)
\begin{align}
\propto& \sum_{\boldsymbol{p}\,\in P}\left( \prod\limits_{r=1}^R\sum_{j=1}^{L_{B,r}}\,\delta_{b_{r,j}}(M_r(\boldsymbol{x};\boldsymbol{a}_{\boldsymbol{p}})) \right) = \sum_{\boldsymbol{p}\,\in P}\sum\limits_{\boldsymbol{q}\,\in Q} \prod\limits_{r=1}^R  \delta_{b_{r,q_r}}(M_r(\boldsymbol{x};\boldsymbol{a}_{\boldsymbol{p}})) \;,\label{equ:fullyexpanded_diracLikelihood}
\end{align}
with all $\#{}Q:=\prod\limits_{r=1}^{R}L_{B,r}$ combinations of the index tuples $\boldsymbol{b}_{\boldsymbol{q}}:=(b_{1,q_1},..,b_{R,q_R})\in Q$ and $q_r\in\{1,..,L_{B,r}\}$, with $\boldsymbol{b}_{\boldsymbol{q}}$ representing one of the $\#{}Q$ combinations of the discrete values $b_{r,j}$ in each entry of $\boldsymbol{b}$. Non--zero probability densities are only at
\begin{align*}
\boldsymbol{M}(\boldsymbol{x};\boldsymbol{a}_{\boldsymbol{p}}) = \boldsymbol{b}_{\boldsymbol{q}}\quad\forall\,\boldsymbol{p}\,\in P,\,\boldsymbol{q}\,\in Q\;.
\end{align*}
This means we are looking for solutions $\boldsymbol{x}$ of $\#{}P\cdot \#{}Q = \left(\prod\limits_{i=1}^{K}L_{A,i}\right)\cdot \left( \prod\limits_{r=1}^{R} L_{B,r}\right)$ discrete equation systems containing all possible permutations of the discrete Dirac components. For demonstration purposes, if we have $R=2$ equations and $L_A=3$ discrete values for all $a_{i}\in\mathbb{R}^8$ and $L_B=5$ for $B_r$, we get $L_A^K\cdot L_B^R = 3^8\cdot 5^2 = 164025$ different equation systems.\medskip

For a deeper understanding, we will introduce a restriction of this general problem description by separating the parameter vector $\boldsymbol{A}$ into a partition $\boldsymbol{A}_r$ ($r=1,..,R$) where only parameters $\boldsymbol{A}_r\in\mathbb{R}^{K_r}$ appear in equation $r$ (such as for linear equation systems, as presented in Subsection \ref{sec:randomLES}), i.e.
\begin{align*}
M(\boldsymbol{x};\boldsymbol{A}_r) &= B_r \quad\forall r=1,..,R\;.
\end{align*}
This represents a restriction, since now random variables $A_i$ cannot appear simultaneously in different equations. {This partitioning does not change the mutual stochastic independency of all $A_i$ ($i=1,..,K$).} Further we assume the same equation type for each equation, i.e. $M_r = M$. This leads now to the total number $K=\sum\limits_{r=1}^{R}K_r$ and, in consequence, to $\prod\limits_{r=1}^{R} \left(\prod\limits_{i=1}^{K_r}L_{A,r,i}\right)\cdot L_{B,r}$ different discrete equation systems. It has to be noted that (if for each given $i\in\{1,..,K_r\}$ redundancy of the mixture model components in the random variables $A_{r,i}$ for all $r=1,..,R$ is excluded) this large number of equation systems contain {$\sum\limits_{r=1}^{R}\left(\prod\limits_{i=1}^{K_r}L_{A,r,i} \right)\cdot L_{B,r} $} different equations (considering only the modes of the densities of $A_{r,i}$ for all $r=1,\dots,R$). For demonstration purposes, if we have $R=2$ equations and $L_A=3$ discrete values for all $a_{r,i}\in\mathbb{R}^4$ and $L_B=5$ for $B_r$ modeled by these Dirac distributions, we would get again $(L_A^K\cdot L_B)^R= (3^{4}\cdot 5)^2 = 164025$ different equation systems (containing {$R\cdot (L_A^K\cdot L_B)=2\cdot 3^4\cdot 5 = 810$} different equations) that $\boldsymbol{x}\in\mathbb{R}^4$ would need to solve.\medskip

This investigation is helpful in a theoretical perspective: 
First, we see the combinatorial extent of equation systems that are simultaneously solved, and this is all captured by a single definition of Equation (\ref{equ:genequsys}) only by using mixture model densities. 
Second, although a simultaneous solution of all such deterministic equation systems seems practically only possible in very specific settings, this changes if we are not using Dirac distributions but extended mixture model component densities, e.g. Gaussian Mixture Models in the context of the most likely solutions. Then the posterior with highest density values shows the values $\boldsymbol{x}$ that fulfill all of the combinations of these equation systems at best.  
Third, Equation (\ref{equ:fullyexpanded_diracLikelihood}) essentially resembles the logic behind what the likelihood function calculates and how using more mixture model components with less equations can be compared to less mixture model components with more equations. Essentially, the likelihood function shows \textit{the sum over all mixture model component permutations of the individual equation system likelihoods}.

\subsection{Random Linear Equation Systems}
\label{sec:randomLES}

Although we have derived the likelihood and posterior for general nonlinear random equation systems, it is instructive to investigate random linear equation systems and we can identify them directly when introducing the definition $M(\boldsymbol{x};\boldsymbol{A}_r) = \boldsymbol{A}_r^T\cdot \boldsymbol{x}$ for $r=1,..,R$ and this can be summarized to
\begin{align*}
A\cdot\boldsymbol{x}=\boldsymbol{B}
\end{align*}
with $A$ of size $R\times n$ with $\boldsymbol{A}_r^T$ in its rows and $\boldsymbol{B}$ with $R$ entries $B_r$, i.e. all parameters are random variables. Such systems are widely applied and under current investigation for specific solution properties, such as nonnegativity of the solution vector $\boldsymbol{x}$ \cite{Landmann2020}. The posterior for this model specification is in the notation of this work ($K=R\cdot n$)
\begin{align}
\pi(\boldsymbol{x}\,\vert\,\boldsymbol{0})\propto& \left( \;\int\limits_{\mathbb{R}^{R\cdot n}} \left(\prod\limits_{r=1}^R\; f_{B_r}(\boldsymbol{s}_r^{\,T}\cdot\boldsymbol{x})\right)\cdot \left(\prod\limits_{i=1}^{K}\;f_{A_{i}}(s_i) \right)\;\text{d}\boldsymbol{s}\right)\cdot \pi(\boldsymbol{x})\label{equ:posterior_LinEquSys_1}\;,
\end{align}
with a linear index $i$ going through the entries of matrix $A$. Utilizing this framework with mixture model random variables, one can determine solutions $\boldsymbol{x}$ in the posterior that fulfill all combinations of equation systems at best. This will be further investigated in the simulation results in Subsection \ref{sec:res:randomlinear}. As a reminder, by defining $A_{i}\sim\delta_{a_{i}}$ and $B_{r}\sim\mathcal{N}(b_{r},\sigma^2)$, i.e. both \textit{not mixture models}, and neglecting positive constants, we get 
\begin{align}
\mathcal{L}(\boldsymbol{0}\,\vert\,\boldsymbol{x})\propto&\prod\limits_{r=1}^R\;  \text{e}^{\frac{1}{2\,\sigma^2}\left(\boldsymbol{a}_r^T\cdot\boldsymbol{x}-b_r\right)^2} = \text{e}^{\frac{1}{2\,\sigma^2}\sum\limits_{r=1}^R \left(\boldsymbol{a}_r^T\cdot\boldsymbol{x}-b_r\right)^2}\label{equ:posterior_LSQ}\;,
\end{align}
which maximum is identical to the classical \textit{least squares} solution.

\subsection{Interpretations and Extensions}
\label{sec:extensions}

\subsubsection{Equation with no Separable Parameter Random Variable}
\label{sec:implicit}

In applications where Equation (\ref{equ:genequsys}) cannot be formulated, since there is no separable $\boldsymbol{B}$, but the equation is defined implicitly with respect to the parameters only by parameters $\boldsymbol{A}$, i.e. 
\begin{align*}
\boldsymbol{M}(\boldsymbol{x};\boldsymbol{A}) &= \boldsymbol{0} \;,
\end{align*}
we propose to utilize the same argumentation as before by introducing artificial mono--modal random variables $B_r$ with the mode $0$ in the entries of $\boldsymbol{B}$ and with standard deviation $\varepsilon$. Utilizing this approach, the technical derivations of Equation (\ref{equ:Likelihood_2}), stay the same and afterwards $\varepsilon$ can tend to $0$ to find best approximate solutions (avoiding the interpretation of $B_r\sim\delta$ leading to $f_{B_r}(M_r(\boldsymbol{x};\boldsymbol{s}))=\delta(M_r(\boldsymbol{x};\boldsymbol{s}))$ for arbitrary (possibly nonlinear) functions $M_r$ in Equation (\ref{equ:Likelihood_2})).

\subsubsection{Optimization Problems}
\label{sec:opt}

If the starting point is an optimization problem with mixture model parameter random variables $\boldsymbol{A}\in\mathbb{R}^K$
\begin{align*}
\text{argmin}_{\boldsymbol{x}}\; F(\boldsymbol{x};\boldsymbol{A})\;, 
\end{align*}
with $F$ a partially differentiable functional $\mathbb{R}^n\rightarrow \mathbb{R}$ needed to be minimized, one can map this to the previous presentation by the necessary condition
\begin{align*}
\nabla_{\boldsymbol{x}}\;F(\boldsymbol{x};\boldsymbol{A})=\boldsymbol{0} \quad\Leftrightarrow\quad \frac{\partial }{\partial x_r} F(\boldsymbol{x};\boldsymbol{A})=0 \quad\forall r=1,..,n\;,
\end{align*}
by interpretation of $M_r(\boldsymbol{x};\boldsymbol{A}) := \frac{\partial }{\partial x_r} F(\boldsymbol{x};\boldsymbol{A})$ and the interpretation of Subsection \ref{sec:implicit}. In order to further achieve a sufficient condition  for optimization, the \textit{Hessian matrix} $H_{F}(\boldsymbol{x};\boldsymbol{A})$ can be calculated for an appropriate functional $F$, which is a \textit{random matrix}, containing mixture model densities in each entry. By investigating the eigenvalue distribution (as typically done in \textit{random matrix theory}, e.g. see Subsection \ref{sec:res:matrix}) one can find probabilistic statements about definiteness of the \textit{random Hessian matrix} and, in consequence, about the existence of local maxima, minima or saddle points. 
{Practical applications can be optimal decision making under uncertainties, such as minimizing robustly the costs of a product including (e.g. mixture model) uncertainties about the component prices.}

\subsubsection{Utilization of the Prior}
\label{sec:UsePrior}

As previously introduced, we only utilize a weakly informative prior $\pi(\boldsymbol{x})=$\textit{const.} since we are not interested in finding best solutions that fit essentially a predefined prior, but want to learn solutions from the equation system. We want to introduce classes of priors that realize hard constraints about the solution space, which can be regarded as part of the problem definition without emphasizing subjective information.\medskip

First, we want to point out that all classical hard and soft constraints about the solution space can be realized by priors, for example, problems where the domain of the solution space is restricted, such as
\begin{align*}
M_r(\boldsymbol{x};\boldsymbol{A}) &= B_r \quad\forall r=1,..,R \quad\text{ s.t. } \lVert\boldsymbol{x}-\boldsymbol{w}\rVert < C\;,
\end{align*}
allowing only solutions in a circle with center $\boldsymbol{w}$ and radius $C$, by introducing a prior \cite{Hoegele_2013}
\begin{align*}
\pi(\boldsymbol{x}) := \left\lbrace\begin{array}{ll} \frac{1}{C^2\,\pi} & \lVert\boldsymbol{x}-\boldsymbol{w}\rVert < C \\ 0 & \text{else} \end{array}\right.\;.
\end{align*}

Second, if the solution space itself is discrete, for example, $\boldsymbol{x}$ on a finite grid then this can be realized by constructing the prior itself as a mixture model. For example, an integer grid
\begin{align*}
M_r(\boldsymbol{x};\boldsymbol{A}) &= B_r \quad\forall r=1,..,R \quad\text{ s.t. } \boldsymbol{x}\in\{-C,..,C\}^n\;,
\end{align*}
can be realized by a prior with
\begin{align*}
\pi(\boldsymbol{x}) :=  \frac{1}{(2\,C+1)^n} \sum\limits_{p=1}^P \delta(\boldsymbol{x}-\boldsymbol{w}_p)\;,
\end{align*}
with $\boldsymbol{w}_p$ all $P$ permutations for all entries from the finite set $\{-C,..,C\}$. Of course, not only the Dirac distributions $\delta$ can be utilized, but also finite approximations. This shows the use of the Bayesian framework for discrete equation systems containing all combinations of parameters on a finite set as well as a discrete solution space.

\subsection{Numerical Remarks}
\label{sec:Numerics}

Although, the combinatorial possibilities underlying the solution of random equation systems with mixture models might be very high, practically the numerical solution is not scaling in runtime, since utilizing Equation (\ref{equ:Likelihood_3}), one always needs to evaluate the same integral. The increase in computational cost results only from the evaluation of a mixture model density with more components compared to a mixture model density with less, which is small. This has the practical consequence, that even high combinatorial possibilities contained in the random equation are straightforwardly handled as demonstrated in the results section.\medskip

Further, for high--dimensional equations classical integration algorithms might be computationally expensive. As one way around this, one can use \textit{Monte Carlo} integration, by sampling $N$ realizations $\boldsymbol{s}_{n}\in\mathbb{R}^{K}$ from the density function $f_{\boldsymbol{A}}$ (which is further simplified for independent entries $A_i$, since only each $f_{A_i}$ needs to be sampled). The integration in the likelihood function of Equation (\ref{equ:Likelihood_2}) can be reduced to
\begin{align}
{\mathcal{L}(\boldsymbol{0}\,\vert\,\boldsymbol{x})\approx \;  \frac{1}{N} \sum\limits_{n=1}^N\left(\prod\limits_{r=1}^R f_{B_r}(M_r(\boldsymbol{x};\boldsymbol{s}_{n}))\right)}\;.\label{equ:MonteCarloInt}
\end{align}

{A major speed-up occurs if the problem is defined as it is typically done in our simulations (except in the portfolio example in Subsection \ref{sec:res:portfolio} and the control engineering in Subsection \ref{sec:res:control}) and discussed already in Subsection \ref{sec:dirac}: We further assume the $K$-dimensional random vector $\boldsymbol{A}$  to be constructed from parameter subvectors $\boldsymbol{A}_r$ (=partitions) with $K_r$  entries (i.e. $K=R\cdot K_r$) each appearing exclusively in the corresponding equation $r=1,..,R$. The advantage is that we then need only to sample these subvectors $\boldsymbol{A}_r$ with $\boldsymbol{s}_{r,n}\in\mathbb{R}^{K_r}$ and the overall integral of a product in Equation (\ref{equ:Likelihood_2}) can be rearranged to a product over indepedent integrals which can be calculated by \textit{Monte Carlo} integration with $N_r$ samples each by the approximation
\begin{align*}
\mathcal{L}(\boldsymbol{0}\,\vert\,\boldsymbol{x})\approx \;  \prod\limits_{r=1}^R\left(\frac{1}{N_r} \sum\limits_{n=1}^{N_r} f_{B_r}(M_r(\boldsymbol{x};\boldsymbol{s}_{r,n}))\right)\;.
\end{align*}
}

For the simulations in the results section we utilized \textit{latin hypercube sampling} for the samples $\boldsymbol{s}_{n}$ leading to a very fast and efficient calculation. {It is worth mentioning that also other numerical strategies not based on \textit{Monte Carlo} integration can be used to evaluate the integral in Equation (\ref{equ:Likelihood_2}). Depending on $K$ (the dimensionality of the integral) and the shape of the integrand, other standard approaches, such as \textit{adaptive quadrature} rules can be utilized efficiently.}\medskip

The only expensive evaluation in this formula is obviously $f_{B_r}$ which can be avoided if the argument $M_r(\boldsymbol{x};\boldsymbol{s}_{n})$ is not in the (practical) support region of $f_{B_r}>0$, which accelerates the evaluation significantly. This can be regarded as the ``pinhole'' which resembles the equation sign in each random equation. \medskip

For a more systematic investigation of computational trade--offs we are following this argumentation: In Equation (\ref{equ:MonteCarloInt}) we can see that the number of evaluations of $f_{B_r}(M_r(\boldsymbol{x};\cdot))$ need to minimized. In the following we assume that we have the same number of mixture model components for all $A_{i}$ and $B_{r}$, denoted by $L$, and the dimension of $\boldsymbol{A}$ is $K$. If we assume that for every mixture model component a fixed number of evaluations $S$ is needed to sample it well, we get $K\cdot L\cdot S$ evaluations (note, one sample with dimension $K$ already samples for all $R$ equations). Further, we have seen that the number of equation system combinations in this setup is $L^{R+K}$. 

We again assume now that the number of $A_i$s, namely $K$, is constructed by $K=R\cdot K_r$, with $K_r$ the constant number of $A_i$ in each equation $r=1,..,R$ (such as in linear equation systems). One question that is of interest is: If we want to investigate a given number $T$ of equation system combinations, what is a good selection for $L$ (number of mixture model components) and $R$ (number of equations in the equation system) to reduce the overall evaluations? By setting $L^{R+K} = L^{R\,(K_r+1)} = T$, we get (for $T,L>1$) $R=\frac{\ln(T)}{(K_r+1)\,\ln(L)}$ and get the total number of evaluations depending on $L$ with $K\cdot L\cdot S = R\cdot K_r\cdot L\cdot S = \frac{K_r\,L\,S\,\ln(T)\,}{(K_r+1)\,\ln(L)}$. If we further ignore constant scaling factors of the desired evaluation $T,S,K_r$, we get a qualitative behavior of $\frac{L}{\ln(L)}$ which has always the discrete minimum at $L=3$ and a comparably slow increase for $L>3$. It is remarkable that the position of this minimum at $L=3$ is independent of all the other problem specific constants (in this argumentation framework). This suggests that using $L=3$ is a reasonable choice but the evaluation times do not strongly increase, for example, for $L=2$ ($+5.6\%$), $L=4$ ($+5.6\%$), $L=5$ ($+13.7\%$) or $L=6$ ($+22.6\%$), except for $L\gg 3$, such as for $L=15$ ($+102.8\%$) essentially doubling the evaluation time. This leads to a rule of thumb of keeping $L\leq 10$ allowing some degree of freedom for modeling the equation systems for efficient evaluations.\medskip

\section{Simulation Results}
\label{sec:results}

The purpose of the results section is to visualize the presented likelihood functions and/or posterior densities (utilizing appropriate priors) by numerical examples in order to improve understanding, meaning and application of the derived formulas. Specifically, we want to demonstrate that the underlying combinations of the modes of the mixture models with practically disjoint components can be handled efficiently by the presented calculation of the likelihood functions. {Please be aware that while the intensity maps of the likelihood functions and posterior densities contain limited information in terms of absolute values, their values are useful for relative comparison within each simulation scenario.}
 
The results section is organized as follows: In Subsection \ref{sec:res:randomlinear} examples for random linear equation systems are presented in order to allow the reader to establish an intuitive connection between deterministic and random equations in this framework. In Subsection \ref{sec:res:randomconic} this intuition is extended to nonlinear system of equations and extreme numbers of mixture model component combinations in order to evaluate the combinatorial potential. In Subsections \ref{sec:res:portfolio}, \ref{sec:res:control} and \ref{sec:res:matrix} it is demonstrated with small simulation examples how this type of modeling can be efficiently utilized for three different application fields. Besides the topical structure of the results section we state \textit{observations \#{}1--9} appearing in the simulation examples in order to point to relevant aspects for this type of modeling and simulation.

\subsection{Demonstration of Solving Random Linear Equation Systems}
\label{sec:res:randomlinear}

\begin{figure}[htbp]
\centering
\includegraphics[width=12cm]{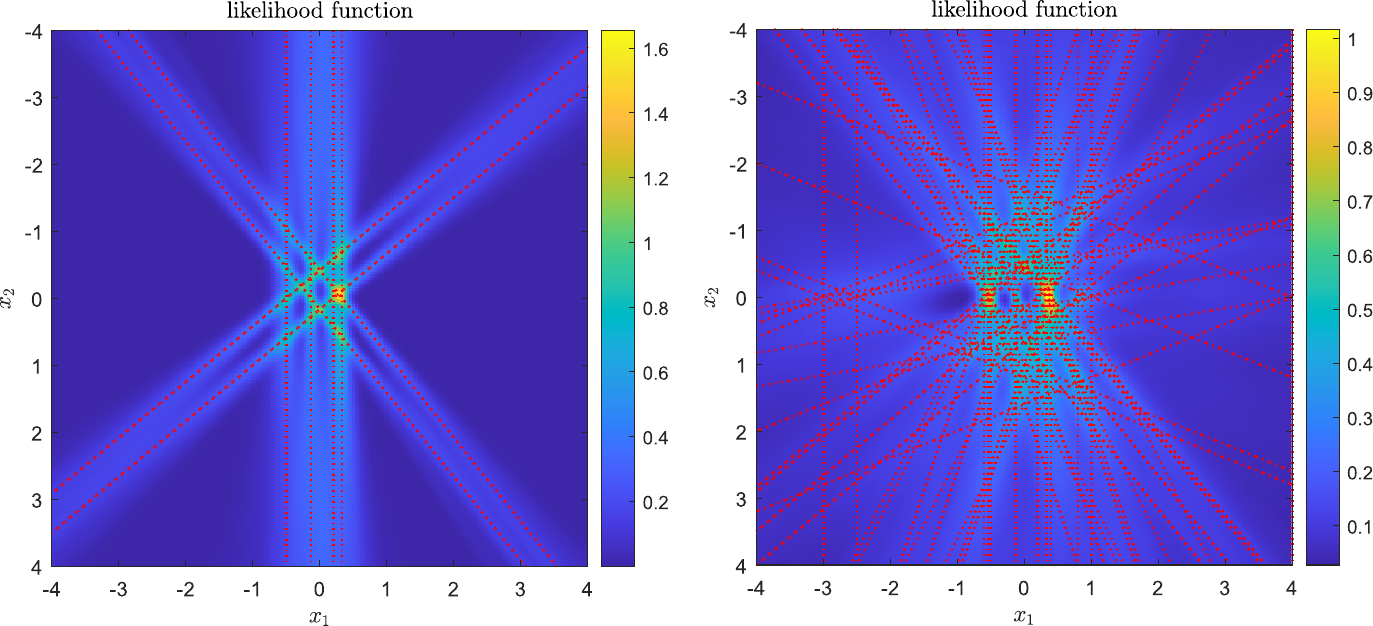}
\caption{Illustration of the solution space of a single random linear equation by showing the exact solutions for all $2^3=8$ (left) and $4^3=64$ (right) combinations of the modes of the Gaussian mixture model parameter densities (red dotted lines) as well as the likelihood function as intensity map.}  
\label{fig:Res:LinearEquation_007}
\end{figure}

For a first demonstration of a random equation system, we are investigating a single random linear equation
\begin{align}
A_{1}\,x_1 + A_{2}\,x_2 &= B
\end{align}
with independent mixture model random variables $A_{1},A_{2},B$. This obviously is an underdetermined system, which would lead to an infinite number of solutions for the deterministic equation. The corresponding likelihood function for the solution $\boldsymbol{x}$ is given by
\begin{align*}
\mathcal{L}(\boldsymbol{0}\,\vert\,\boldsymbol{x})\propto\int\limits_{\mathbb{R}^2} f_{B}(s_1\,x_1+s_2\,x_2)\cdot f_{A_{1}}(s_1)\cdot f_{A_{2}}(s_2)\;\text{d}\boldsymbol{s}\;.
\end{align*}
For simplicity, we define every random variable in this equation as Gaussian Mixture Models (GMMs) which are composed of separated Gaussians components in order to allow simultaneously different possibilities for the parameters. In Figure \ref{fig:Res:LinearEquation_007} the likelihood function (=solution space) for example equations with $L=2$ Gaussian components (left) and with $L=4$ Gaussian components (right) are presented. As red dotted lines, the solutions of all combinatorial possibilities of the modes of the Gaussians are presented and the likelihood function as intensity map. One can clearly see that the likelihood function is aligning very well to the exact mode solutions. The overall highest intensities of these likelihood functions represent the solutions $\boldsymbol{x}$ that fulfill the equation system with all parameter mixture models at best.

\begin{figure}[htbp]
\centering
\includegraphics[width=12cm]{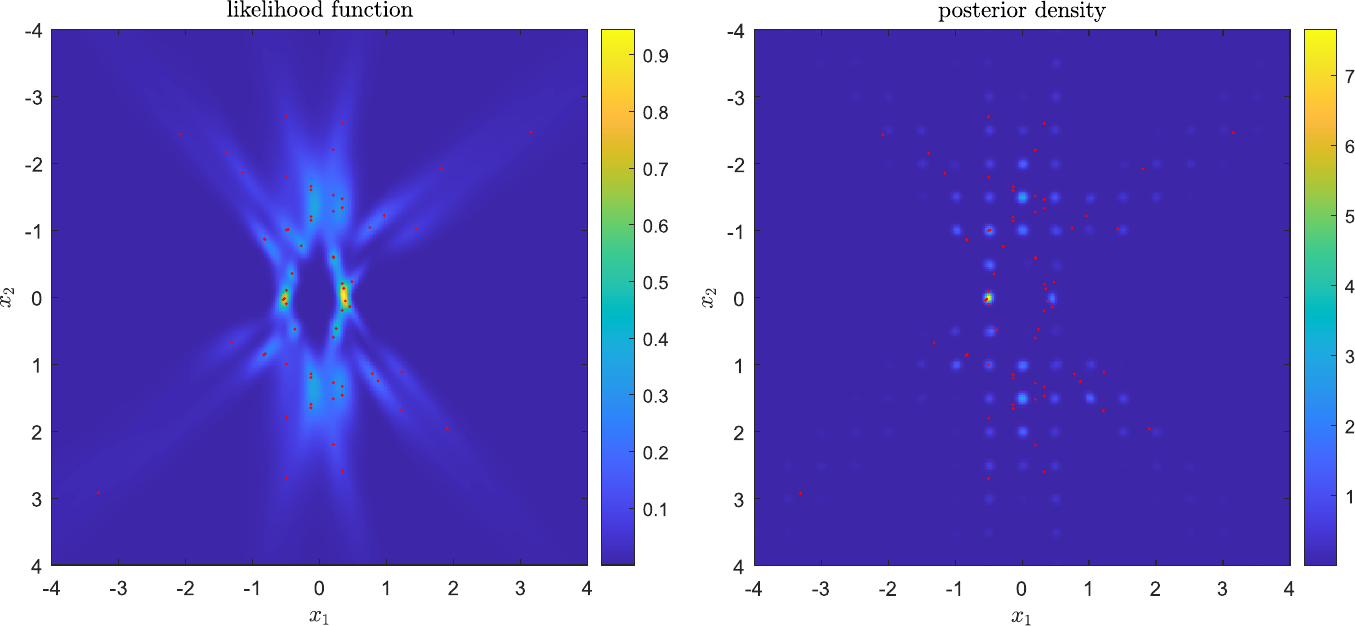}
\caption{Illustration of the solution space of a $2\times 2$ random linear equation system by showing the exact solutions for all $64$ combinations of the modes of the Gaussian mixture model parameter densities (red points). Left and right show the same random equation system with the same standard deviation of the Gaussians $\sigma = 0.1$. Left: The likelihood function as intensity map. Right: The posterior density with a prior describing a Gaussian grid.}  
\label{fig:Res:LinearEquation_001}
\end{figure}

Next, we want to explicitly show the solution space of a simple $2\times 2$ linear equation system $A\cdot\boldsymbol{x} = \boldsymbol{B}$
\begin{align*}
A_{1,1}\,x_1 + A_{1,2}\,x_2 &= B_1\\
A_{2,1}\,x_1 + A_{2,2}\,x_2 &= B_2
\end{align*}
with independent mixture model random variables $A_{i,j}$ and $B_{i}$ ($i,j\in\{1,2\}$) (note, we utilize double indexing as typical for matrices, but in the general theory the $A_{i,j}$ can be regarded as a linear index running from $A_1,..,A_4$ as, for example, in Equation (\ref{equ:posterior_LinEquSys_1})). We are again utilizing a GMM with $L=2$ components. This corresponds inherently to $2^6=64$ different combinations of mono--modal stochastically blurred linear equations. In Figure \ref{fig:Res:LinearEquation_001} the exact solutions for all $64$ combinations of the modes of the Gaussians are presented as red dots as well as the likelihood function in an intensity map. It can be directly seen that these likelihood functions capture the solutions of all mode combinations as well as the additional uncertainties from the Gaussian components. For the generation of Figure \ref{fig:Res:LinearEquation_001} (left) (two equations) and Figure \ref{fig:Res:LinearEquation_007} (right) (single equation) we utilized the same Gaussian components but for the $2\times 2$ case separated into two independent equations for the corresponding parameter random variables. One can see that this makes a significant difference for the likelihood function with much more distinct maxima in the $2\times 2$ case. The reason comes from the meaning of the equation systems: In the single equation case, we see the likelihood of the combination of all underdetermined solutions (straight lines) and in the $2\times 2$ case we see the likelihood of all combinations of intersection points, which already utilizes the information of intersection.

\textit{Observation \#{}1:} There are different possibilities to model the same approximate solutions of random equation systems with mixture models by adapting the number of equations and the mixture model components accordingly. In such scenarios the corresponding likelihood functions are significantly different but with high intensity regions at the same positions.

In Figure \ref{fig:Res:LinearEquation_001} (right) we demonstrate the use of a prior on a finite grid of Gaussians and present the corresponding posterior density as discussed in Subsection \ref{sec:UsePrior} showing a clear approximate solution on that grid.

\begin{figure}[htbp]
\centering
\includegraphics[width=12cm]{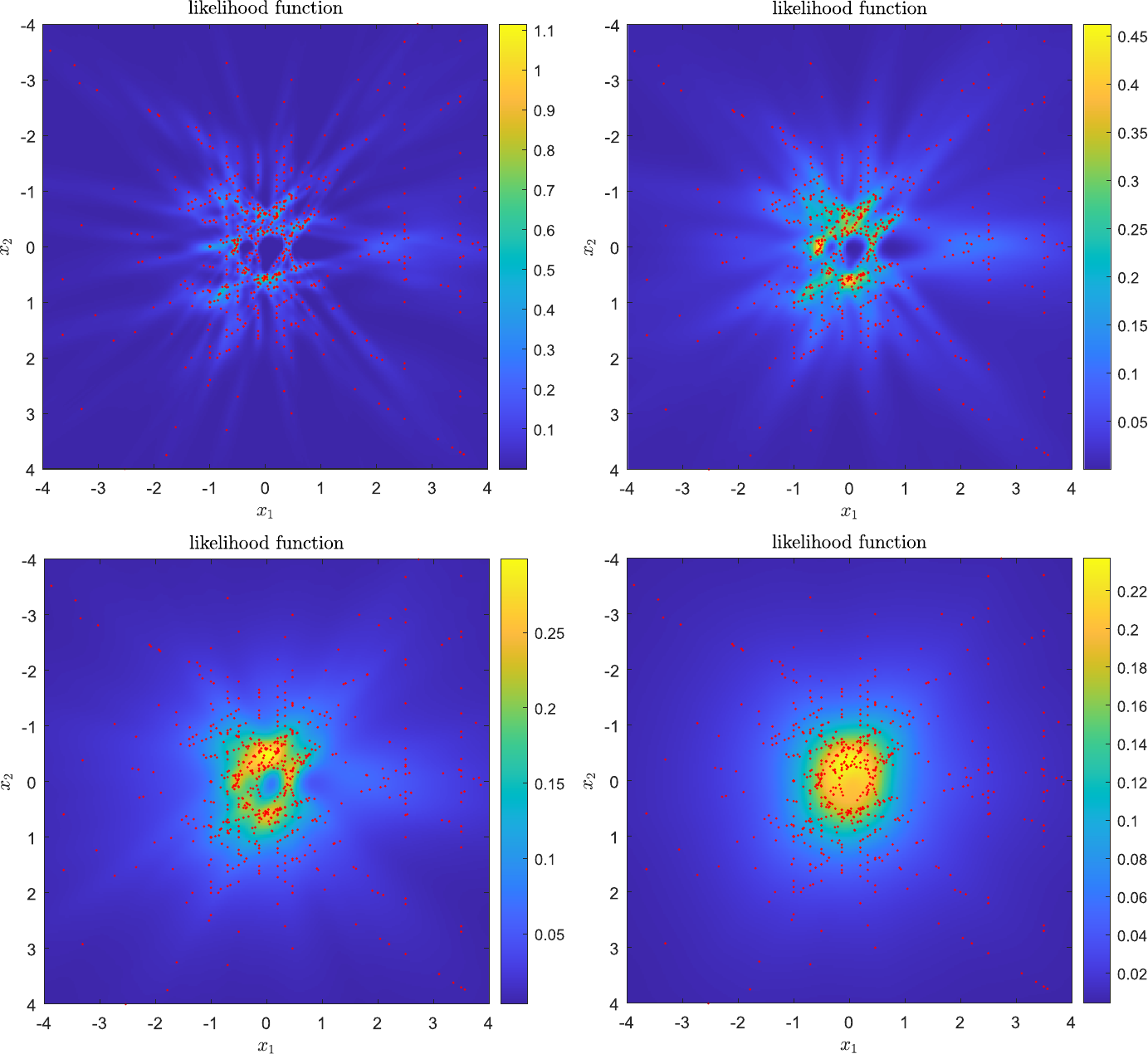}
\caption{Illustration of the solution space of a $2\times 2$ random linear equation system by showing the exact solutions for all $729$ equation systems (red points) as well as the likelihood function as intensity map. Presented is the same random equation system but with increased standard deviations of the individual Gaussians $\sigma = 0.05$ (top left) $\sigma = 0.1$ (top right), $\sigma = 0.2$ (bottom left) and $\sigma = 0.4$ (bottom right).}  
\label{fig:Res:LinearEquation_003}
\end{figure}

In Figure \ref{fig:Res:LinearEquation_003} analog results with Gaussian mixture models with $L=3$ Gaussians for each random variable, resulting in ${3^6=} 729$ equations systems containing {$2\cdot 3^3=54$} different equations, are presented. This shows the combinatorial increase by just adding one Gaussian to each Gaussian mixture model. Four simulation examples of the same random equation system but with increasing standard deviations of the Gaussians are presented. 

\textit{Observation \#{}2:} The position of the highest densities varies significantly and not in a predictable way depending on the standard deviations of the mixture model components. This shows the difficulty to estimate approximate solutions and to find appropriate problem--specific mixture models.

\begin{figure}[htbp]
\centering
\includegraphics[width=12cm]{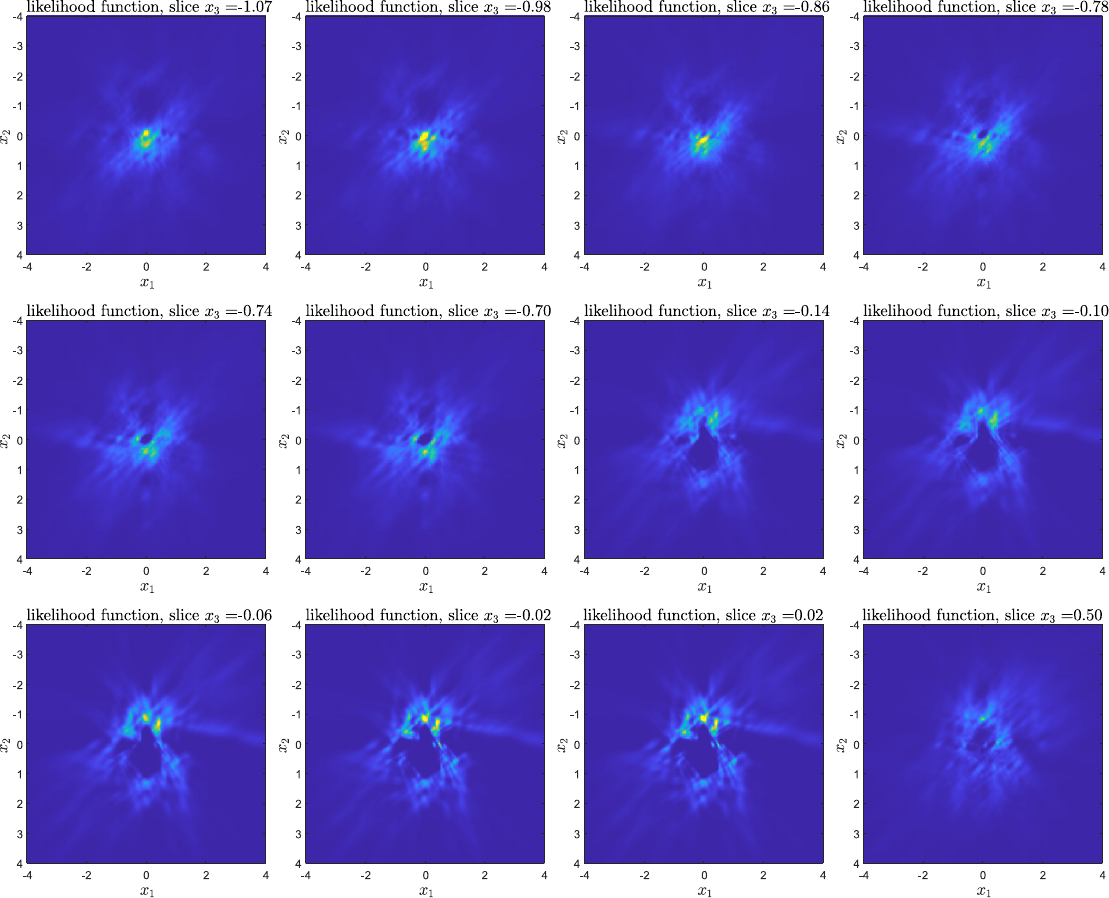}
\caption{Illustration of the solution space of a $3\times 3$ random linear equation system by showing the likelihood function as intensity map. Since $\boldsymbol{x}\in\mathbb{R}^{3}$ the likelihood function is presented in slices of highest intensities with constant $x_3$. Each mixture random variable consists of $L=4$ Gaussian components. {The color scale is constant for all intensity maps in order to allow comparisons.}}  
\label{fig:Res:LinearEquation_011}
\end{figure}

Finally, we investigate the $3\times 3$ linear system
\begin{align*}
A_{1,1}\,x_1 + A_{1,2}\,x_2 + A_{1,3}\,x_3 &= B_1\\
A_{2,1}\,x_1 + A_{2,2}\,x_2 + A_{2,3}\,x_3 &= B_2\\
A_{3,1}\,x_1 + A_{3,2}\,x_2 + A_{3,3}\,x_3 &= B_3
\end{align*}
with independent mixture model random variables $A_{i,j}$ and $B_{i}$ ($i,j\in\{1,2,3\}$). We utilize $L=4$ Gaussian components for each mixture model with $\sigma = 0.12$ leading to $4^{12}$ approx. $1.67\,10^7$ (approx. $16.7$ million) different $3\times 3$ equation systems (containing {$3\cdot 4^4=768$} different equations). The corresponding likelihood function is in $\mathbb{R}^3$ and, in consequence, in Figure \ref{fig:Res:LinearEquation_011} slices of the highest likelihood values for constant $x_3$ are presented to gain an impression of the complexity of the inherent solution space for this $3\times 3$ equation system.

\textit{Observation \#{}3:} Although the mathematical description of a random equation system might have a very low complexity (e.g. $3\times 3$ linear equation system), the solution space can contain highly complex structures due to the high combinatorial potential contained in the utilized mixture models.

\subsection{Demonstration of Solving Random Conic Section Equations}
\label{sec:res:randomconic}

\begin{figure}[htbp]
\centering
\includegraphics[width=12cm]{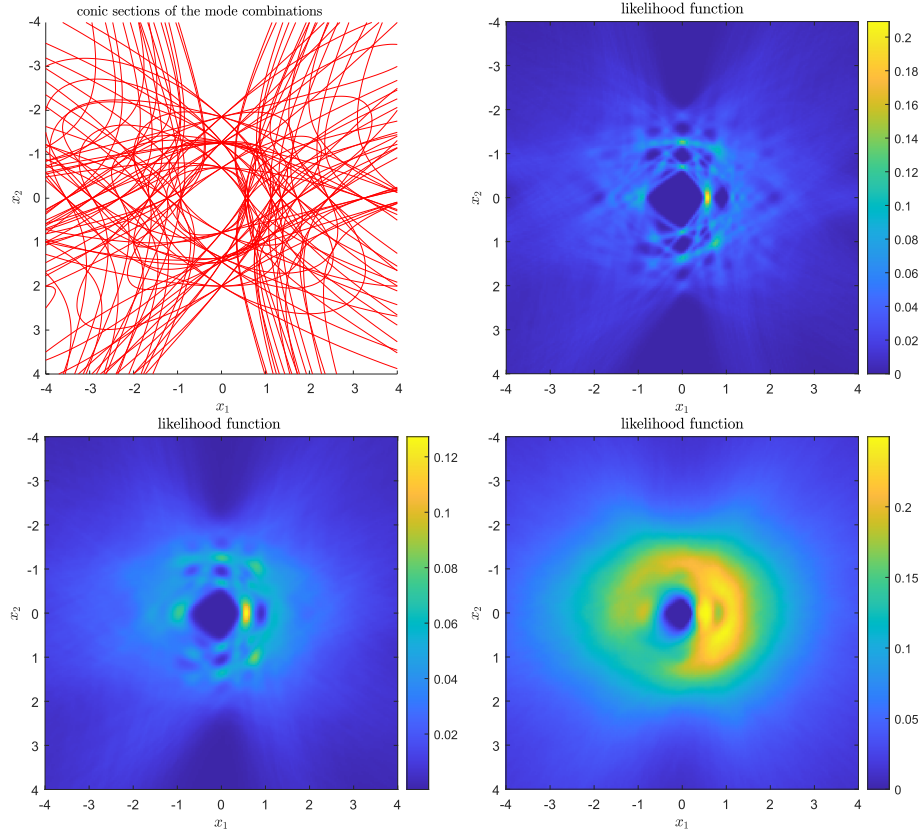}
\caption{Illustration of the solution space of a random conic section equation by showing the conic sections as well as the likelihood function as intensity map. Top left: The implicitly defined conic sections of the exact modes of the mixture models (red lines). There are $L=2$ components utilized in each parameter mixture model and the components are essentially the same with differences described as follows: Top right: Gaussian components with $\sigma = 0.3$. Bottom left: Gaussian components with $\sigma = 0.6$. Bottom right: Uniform components with $\sigma = 0.6$.}  
\label{fig:Res:ConicSection_001}
\end{figure}

In this subsection we are presenting nonlinear, implicitly defined functions $M$ by the investigation of the equations of the conic sections
\begin{align*}
A_{1}\cdot x_1^2 + A_{2}\cdot x_1\cdot x_2+ A_{3}\cdot x_2^2 + A_{4}\cdot x_1 + A_{5}\cdot x_2  &= B
\end{align*}
with independent mixture model random variables $A_{j}$ and $B$ ($j\in\{1,2,3,4,5\}$). This time we assume Gaussian and Uniform mixture models in order to present also that the derived formulas and the efficient evaluation are not dependent on the specific type of density function. In Figure \ref{fig:Res:ConicSection_001} three examples are presented with $L=2$ components of the Gaussians leading to $2^6=64$ mode combinations of conic sections. In top right and bottom left GMMs are utilized with a doubling of the standard deviation from top right to bottom left. On the bottom right, a uniform mixture model is utilized with the same standard deviation for each component as for the Gaussians in bottom left, showing a clearly different shape than the likelihood function based on Gaussians.

\textit{Observation \#{}4:} The position of highest intensity values of the likelihood varies significantly for different types of mixture model components. This shows the nontrivial dependency of the solution space on the mixture model density type.

\begin{figure}[htbp]
\centering
\includegraphics[width=11cm]{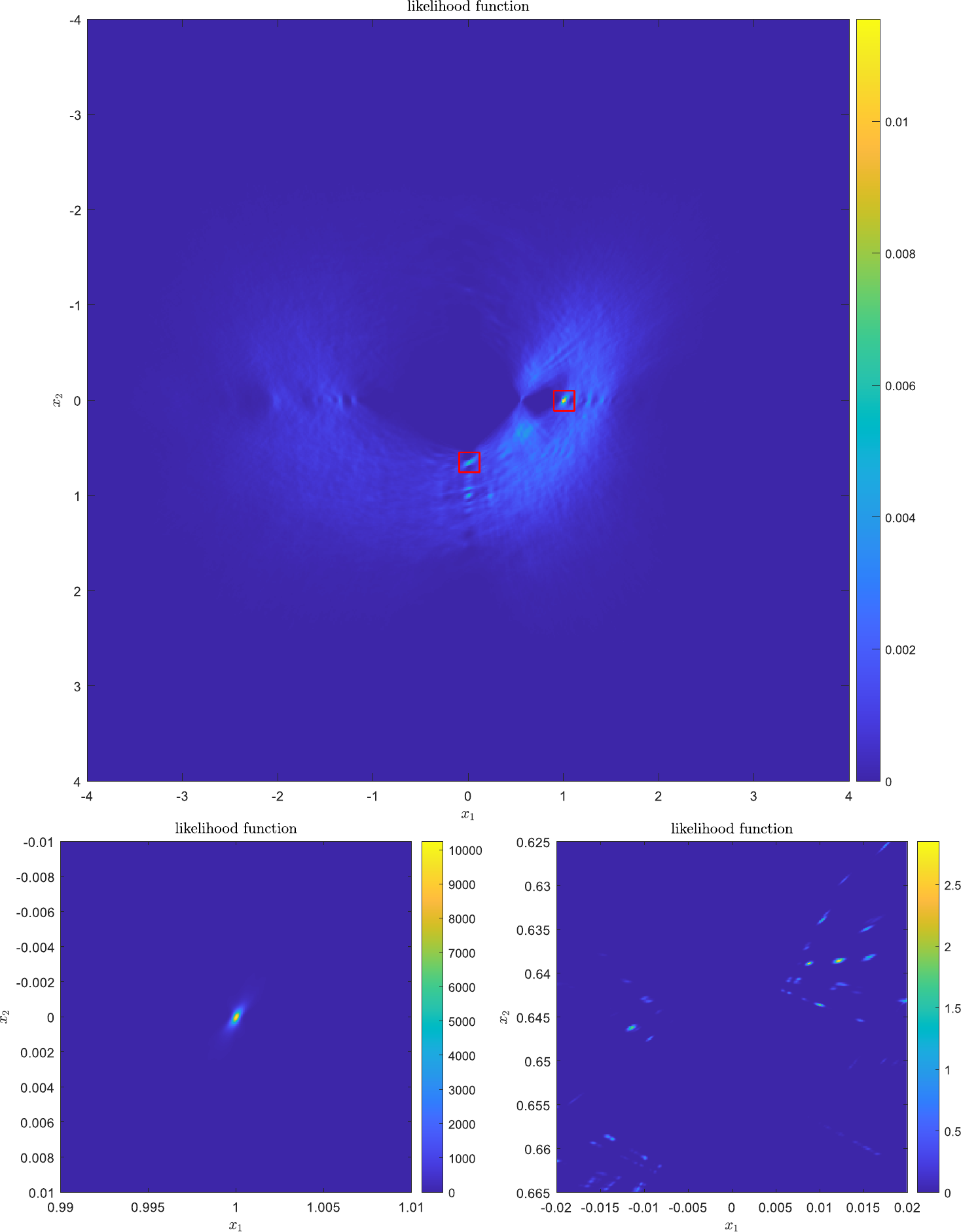}
\caption{Illustration of the solution space of $R=3$ random conic section equations by showing the likelihood function as intensity map. For the calculation of this result roughly $69$ billion different intersections of $3$ conic sections are considered on a $512\times 512$ intensity map. Top: intensity map on a large scale and standard deviation $\sigma =0.1$ of each Gaussian component. Bottom: zoom in two central regions with high intensities at $(1,0)$ (left) and $(0,65)$ (right) with a reduction of the standard deviation to  $\sigma =0.001$ of each Gaussian component, showing different types of local structures.}  
\label{fig:Res:ConicSection_002}
\end{figure}

In order to demonstrate the combinatorial power of this approach we are investigating larger equation systems consisting of $R$ conic section equations
\begin{align*}
A_{r,1}\cdot x_1^2 + A_{r,2}\cdot x_1\cdot x_2+ A_{r,3}\cdot x_2^2 + A_{r,4}\cdot x_1 + A_{r,5}\cdot x_2  &= B_{r}\quad\forall r=1,..,R
\end{align*}
with independent mixture model random variables {$A_{r,j}$ and $B_{r}$ ($r\in\{1,.., R\},j\in\{1,2,3,4,5\}$)} and $R\in\{3,20\}$ (note, for simplicity, we utilize a double indexing of {$A_{r,j}$} again although in the general theory a linear index running form $A_1,..,A_{R\cdot 5}$ is used). Intersections of conic sections are still part of ongoing research \cite{Chomicki_2023}. First, we consider the case of three equations ($R=3$): By assuming for all mixture models $L=4$ components of these $18$ random variables, we get $4^{18}\approx 6.8719\cdot 10^{10}$ (approximately $69$ billion) nonlinear equation systems which correspond to different intersections of $3$ conic sections (containing {$3\cdot 4^6 = 12288$} different conic sections)  that are considered in finding the best solution $\boldsymbol{x}$. In Figure \ref{fig:Res:ConicSection_002} the result of the corresponding likelihood function is presented on a $512\times 512$ grid, clearly indicating the area of high likelihood, which corresponds to the points of most intersections for the given parameter combinations of the conic sections. The calculation time on a standard working laptop with a $1.8$Ghz single core utilizing a standard \textit{Matlab} code took approximately $90$ seconds. By repeating the simulation with differently drawn \textit{Monte Carlo} random numbers, the differences are only on a minor noise level and the structure of the likelihood function is the same. Moreover, as roughly demonstrated in Figure \ref{fig:Res:ConicSection_002} (bottom row), the variation of standard deviations of Gaussians components can be utilized in a search routine to find with singular (bottom left) or clusters (bottom right) of approximate solutions if the width of the Gaussian components itself are not inherent and only auxiliary descriptions of the problem. The logic would be like this: The more one zooms into specific areas of the solution space, the smaller the standard deviations have to be. 

\textit{Observation \#{}5:} By utilizing varying standard deviations of mixture model components in an iterative scheme, one can computationally introduce a {general optimization} search strategy {for approximate equation solutions}, which {can} distinguish between singular and clusters of approximate solutions. 

\begin{figure}[htbp]
\centering
\includegraphics[width=11cm]{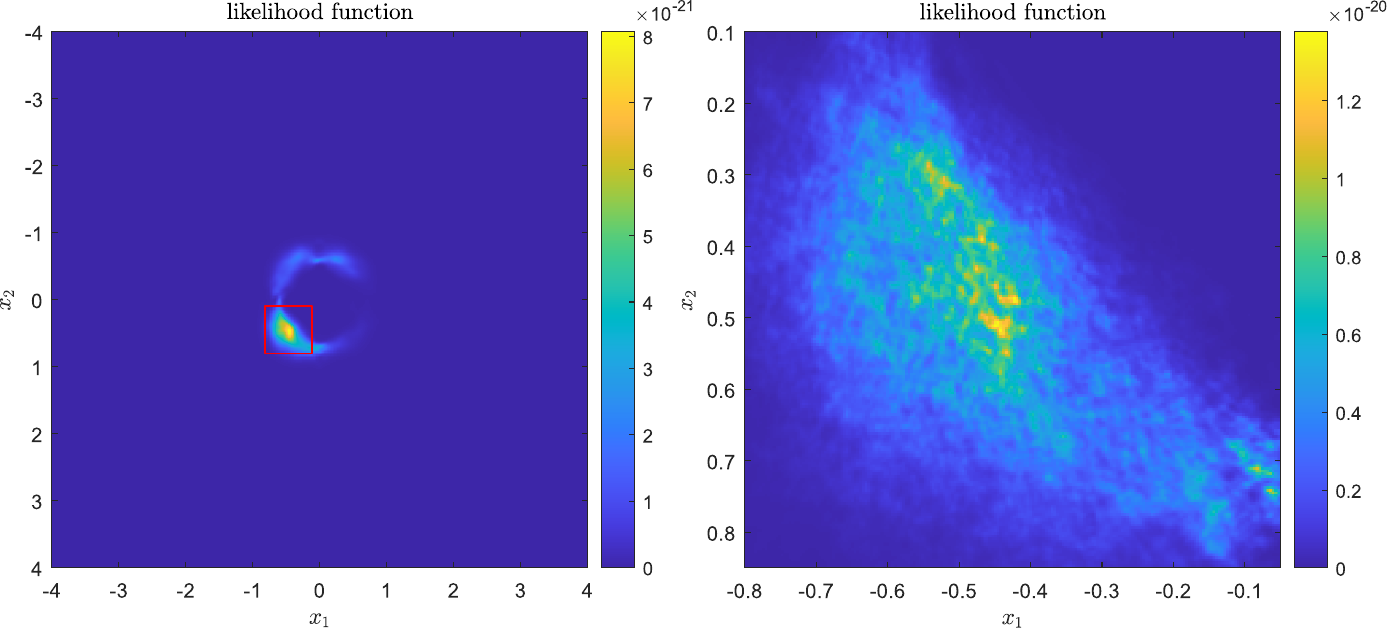}
\caption{Illustration of the solution space of $R=20$ random conic section equations by showing the likelihood function as intensity map. For the calculation of this result roughly $2.4\cdot 10^{93}$ different intersections of $20$ conic sections are considered on two $512\times 512$ intensity maps with standard deviation of each Gaussian component of $\sigma=0.2$ (left) and a zoom into the maximum intensity area $(-0.4,0.5)$ with $\sigma=0.02$ (right).}  
\label{fig:Res:ConicSection_003}
\end{figure}

In Figure \ref{fig:Res:ConicSection_003} we are presenting the most extreme combinatorial simulation of this study by further extending the random equation systems to $R=20$ conic section equations with $L=6$ components of each of the $120$ mixture model random variables leading to $6^{120}\approx 2.4\cdot 10^{93}$ different configurations of intersections of $20$ conic sections (a number larger than the number of estimated particles in the visible universe \cite{Vopson2021}), containing {$20\cdot 6^6 = 933120$} different conic sections. The computation took approximately $1160$ seconds. Again, the differences of reruns are only on a minor noise level with overall the same appearance of the likelihood function.

\textit{Observation \#{}6:} Utilizing \textit{Monte Carlo} integration (Equation (\ref{equ:MonteCarloInt})) is a numerical efficient way to calculate approximate solutions even for equation systems with an extremely high number of combinations. This is based neither on a specifically simple equation type nor on specific mixture model types.

\subsection{Application to Portfolio Optimization}
\label{sec:res:portfolio}

In this application, we want to demonstrate the practical use of the presented methodology, for example, with a straightforward analysis in portfolio optimization. We are assuming to have possible asset returns for $n$ assets $A_1,..,A_n$ as well as the weights to allocate those assets $x_1,..,x_n$ with $x_1,..,x_n>0$ and $\sum\limits_{i=1}^n x_i=1$, leading to the random variable of possible total returns of the portfolio
\begin{align*}
\sum\limits_{i=1}^n x_i\,A_i = \boldsymbol{x}^T\,\boldsymbol{A} \quad \text{ s.t. } x_1,..,x_n>0\;\wedge\;\sum\limits_{i=1}^n x_i=1\;.
\end{align*}
Each random variable $A_i$ contains different possible returns, such as for $A_i>0$ for profits and $A_i<0$ for losses. In the context of this work, every asset return is a mixture model with several separated realization possibilities (and certain deviations from it described by each mixture model component). This can lead to a large number of possible combinations of asset returns.

Further, the random variable of the variance of the total returns for a specific asset allocation can be described by the covariance matrix multiplied by the weights $\boldsymbol{x}$ with
\begin{align*}
\boldsymbol{x}^T\left(\boldsymbol{A}-\mathbb{E}[\boldsymbol{A}]\right)\left(\boldsymbol{A}-\mathbb{E}[\boldsymbol{A}]\right)^T\boldsymbol{x}\;,
\end{align*}
with $\mathbb{E}[\boldsymbol{A}]$ the expectation value vector of the assets. Please note, this is the random variable of the variance of a specific asset allocation and not the variance itself, which would be the expectation value of this random variable.

A main approach in portfolio optimization is to find the weights $\boldsymbol{x}$ in order to maximize the asset return as well as to reduce the variation (reducing the risk). For example, in the \textit{Mean--Variance} framework \cite{Zhang2018} this can be achieved by an optimization problem with a risk adjustment factor $\lambda\in\mathbb{R}^+$
\begin{align*}
&\text{argmax}_{\boldsymbol{x}}\quad \boldsymbol{x}^T\,\boldsymbol{A} - \frac{\lambda}{2}\cdot \boldsymbol{x}^T\left(\boldsymbol{A}-\mathbb{E}[\boldsymbol{A}]\right)\left(\boldsymbol{A}-\mathbb{E}[\boldsymbol{A}]\right)^T\boldsymbol{x}\quad \\
&\text{ s.t. } x_1,..,x_n>0\;\wedge\;\sum\limits_{i=1}^n x_i=1\;.
\end{align*}
Please note, the classical approach would be to minimize the expectation value of this expression \cite{Zhang2018}. An alternative approach presented here is to work with the full densities in order to account better for all the potential outcomes of this objective function as will be demonstrated in the following.
First, we apply the equality constraint by 
\begin{align*}
&\text{argmax}_{\boldsymbol{y}}\quad \boldsymbol{y}^T\,\boldsymbol{A} - \frac{\lambda}{2}\cdot \boldsymbol{y}^T\left(\boldsymbol{A}-\mathbb{E}[\boldsymbol{A}]\right)\left(\boldsymbol{A}-\mathbb{E}[\boldsymbol{A}]\right)^T\boldsymbol{y}\quad \\
&\text{ s.t. } y_1,..,y_{n-1}>0\;\wedge\;\sum\limits_{i=1}^{n-1} y_i<1\;,
\end{align*}
with $\boldsymbol{y} := (x_1,\dots,x_{n-1},1-\sum\limits_{i=1}^{n-1} x_i)^T$. This reduces the number of degrees of freedom by one and replaces the equality constraint by an inequality constraint. By rearranging, this is again a quadratic polynomial in $(x_1,\dots,x_{n-1})$. Second, the necessary condition (along the lines of Subsection \ref{sec:opt}) for this maximization is that the gradient with respect to $\boldsymbol{y}$ of this objective function is zero, which is a random linear equation system with constraints containing $L_A^n$ different mode combinations (assuming $L_A$ mixture model components for each random variable $A_i$). 
For demonstration purposes, we assume a simplified setup: We have just $n=3$ assets with $\mathbb{E}[\boldsymbol{A}]=\boldsymbol{\mu}=(0.2,0.1,0.3)^T$ and compared to these expected returns large mixture model deviations, and each asset return is composed of $6$ Gaussian mixture model components (leading to $6^3=216$ different mode combinations). This leads to the optimization problem for the degrees of freedom $\boldsymbol{x}=(x_1,x_2)^T$ (and the definition $x_3:=1-x_1-x_2$)
\begin{align*}
&\text{argmax}_{\boldsymbol{x}}\quad A_3-\frac{\lambda}{2}\,(A_3-\mu_{3})^2 + \boldsymbol{x}^T\,\left(\begin{array}{c}
(A_1-A_3)-\lambda\,(A_3-\mu_3)\,( (A_1-\mu_1)-(A_3-\mu_3) )\\
(A_2-A_3)-\lambda\,(A_3-\mu_3)\,( (A_2-\mu_2)-(A_3-\mu_3) )
  \end{array}\right)\\
&\hspace{1.5cm}-\frac{\lambda}{2}\cdot \boldsymbol{x}^T\left(\begin{array}{c} (A_1-\mu_1)-(A_3-\mu_3) \\(A_2-\mu_2)-(A_3-\mu_3) \end{array}\right)\cdot\left(\begin{array}{c} (A_1-\mu_1)-(A_3-\mu_3) \\(A_2-\mu_2)-(A_3-\mu_3) \end{array}\right)^T\boldsymbol{x}\\
&\text{ s.t. } x_1,x_2>0\;\wedge\;x_1+x_2<1\;.
\end{align*}
The corresponding necessary condition is the system of random linear equations
\begin{align*}
&\left(\begin{array}{c}
(A_1-A_3)-\lambda\,(A_3-\mu_3)\,( (A_1-\mu_1)-(A_3-\mu_3) )\\
(A_2-A_3)-\lambda\,(A_3-\mu_3)\,( (A_2-\mu_2)-(A_3-\mu_3) )
  \end{array}\right)\\
&\hspace*{1cm} -\lambda\cdot\left(\begin{array}{c} (A_1-\mu_1)-(A_3-\mu_3) \\(A_2-\mu_2)-(A_3-\mu_3) \end{array}\right)\cdot\left(\begin{array}{c} (A_1-\mu_1)-(A_3-\mu_3) \\(A_2-\mu_2)-(A_3-\mu_3) \end{array}\right)^T\boldsymbol{x} = \boldsymbol{0}\\
&\text{ s.t. } x_1,x_2>0\;\wedge\;x_1+x_2<1\;.
\end{align*}
The constraints about the feasibility region of $x_1,x_2$ can be realized according to the presentation in Subsection \ref{sec:UsePrior} by utilizing a suitable prior density.

\begin{figure}[htbp]
\centering
\includegraphics[width=11cm]{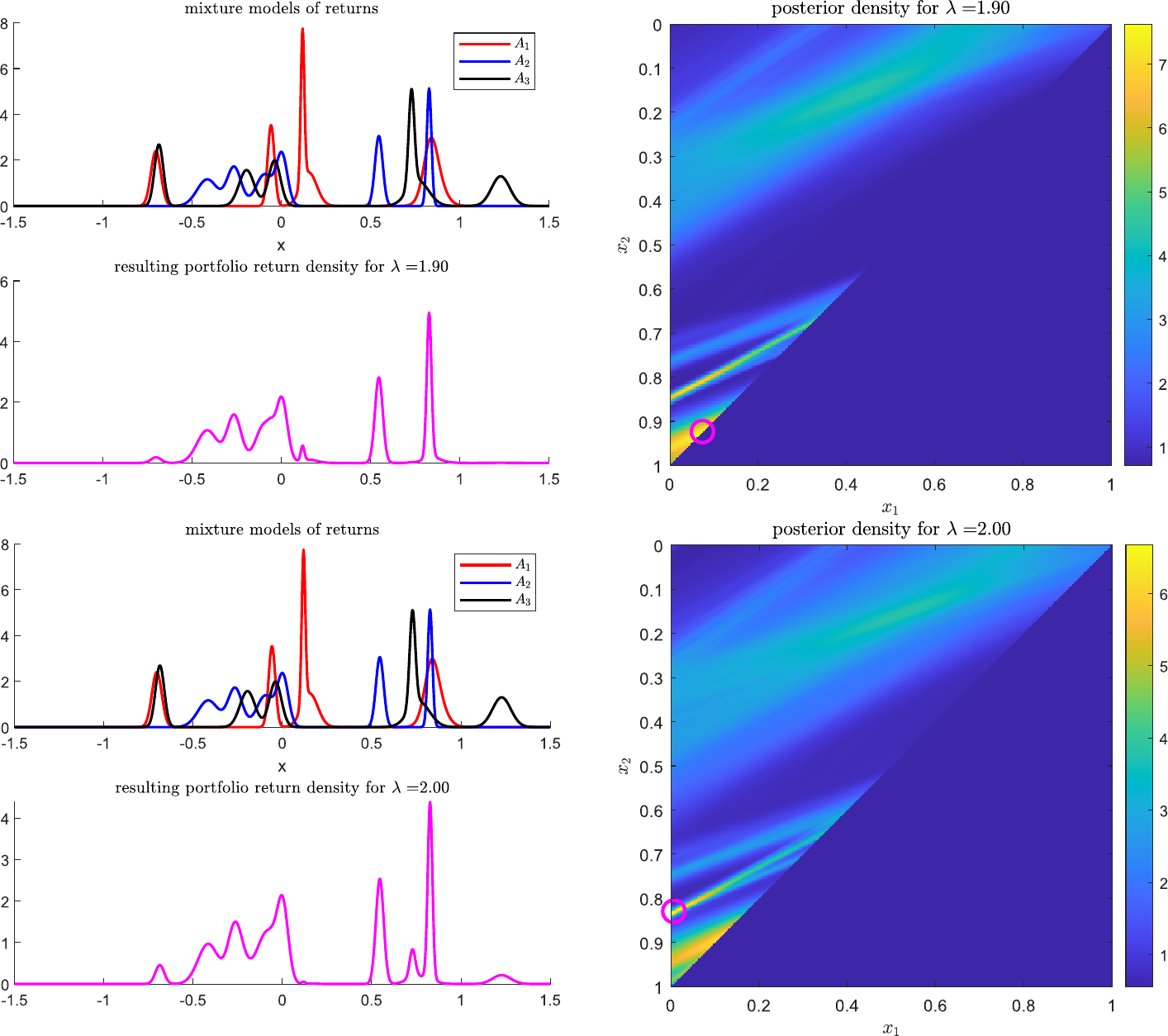}
\caption{Illustration of the portfolio optimization for two scenarios of two slightly different risk adjustment factors $\lambda=1.9$ (top row) to $\lambda=2.0$ (bottom row). In each plot there is presented: Top left: The three mixture model densities of each asset. Right: the posterior density showing the approximate satisfaction of the necessary condition of portfolio optimization as the maximum in this map presented as magenta circle. Bottom left: The resulting return density according to the observed maximum.}  
\label{fig:Res:Portfolio_002}
\end{figure}

In the following, we are investigating this simplified random linear equation system by an experimental simulation. In Figure \ref{fig:Res:Portfolio_002} two scenarios with only a slight change of the risk adjustment factor from $\lambda=1.9$ (top row) to $\lambda=2.0$ (bottom row) are presented. It shows the non--continuous nature of this type of portfolio optimization in the posterior density ({switching discontinuously from the asset combination of $A_1$ and $A_2$ for $\lambda=1.9$ to the combination of $A_2$ and $A_3$ for $\lambda=2.0$}). Please note, such effects would not be visible if we would only have taken the expectation value of these equations and solve it, which is a deterministic linear equation system showing no discontinuities in the solution space with respect to $\lambda$.
For a more detailed portfolio optimization, investigations of the \textit{random Hessian matrix} would be necessary to distinguish between maxima, minima and saddle points of the total return for asymmetric individual asset returns for a large number of possible assets, which is not part of this application example.

\textit{Observation \#{}7:} Random equation systems with mixture models can be used for optimization problems with significant combinatorial components as part of the problem definition, such as the efficient analysis in portfolio management, especially for a large number of non--trivial asset returns with large deviations and high combinatorial potentials.

\subsection{Application to Control Engineering}
\label{sec:res:control}

In this subsection, we want to demonstrate the application of random equation systems to control engineering. Besides a long relation between system descriptions with random variables in control theory \cite{Astroem1970}, there is sophisticated current research on the general embedding of control problems in the space of random variables \cite{Bensoussan2019}. We want to demonstrate the use of mixture models in this context in a simplified example setup for a control problem defined by a difference state equation in discrete time, open to a broad readership. We will illustrate the control task by an example of a chemical heating process accompanying the general formulation. The simplified setup is as follows: There is a current state $\boldsymbol{S}_i$ of a system at time index $i=1,..,n$ (with $\boldsymbol{S}_1$ the initial state) (in the example, this could be the temperature of a chemical). Further, the system has a target state it should reach, denoted by $\boldsymbol{S}_{\text{target}}$ (in the example, this could be the target temperature of the chemical). As last input to the problem definition, the update of the state at time $i$ is described by
\begin{align*}
\Delta \boldsymbol{S}_i = \boldsymbol{G}(\boldsymbol{u}_i) + \boldsymbol{H}(\boldsymbol{S}_i) \;,
\end{align*} 
with $\boldsymbol{G}$ the (possibly nonlinear) function that translates the control action $\boldsymbol{u}_i$ at time index $i$ to the state update (in the example, this could be the adjustable electrical current $\boldsymbol{u}_i$ in a cooling device that leads to a reduction of the temperature in the chemical the higher the current). Further $\boldsymbol{H}$ represents the (possibly nonlinear) inherent update function of the system that dynamically changes the next state depending on the current state $\boldsymbol{S}_i$ (in the example, this could be internal heating of the chemical due to a chemical process depending on the temperature $\boldsymbol{S}_i$). This leads to a control engineering problem if the total updates (control action and inherent update) should reach the target state $\boldsymbol{S}_{\text{target}}$, i.e. leading to the control loop (with a given initial state $\boldsymbol{S}_1$)
\begin{align}
\text{(solve for $\boldsymbol{u}_i$ based on control deviation)} \quad& \boldsymbol{S}_{\text{target}}-\boldsymbol{S}_i = \boldsymbol{G}(\boldsymbol{u}_i) + \boldsymbol{H}(\boldsymbol{S}_i)\label{equ:control1}\\
\text{(update state $\boldsymbol{S}_{i}\rightarrow\boldsymbol{S}_{i+1}$ )} \quad& \boldsymbol{S}_{i+1} = \boldsymbol{S}_i + \boldsymbol{G}(\boldsymbol{u}_i) + \boldsymbol{H}(\boldsymbol{S}_i)\;,\label{equ:control2}
\end{align}
for $i=1,..,n-1$. In this control loop first, the best control action needs to be decided in order to meet the target state $\boldsymbol{S}_{\text{target}}$ (Equation (\ref{equ:control1})) (in the example, to which value the electrical current $\boldsymbol{u}_i$ should be adjusted in the cooling device) and then the system state is updated to $\boldsymbol{S}_{i+1}$ following this action (Equation (\ref{equ:control2})) (in the example, for one time step the external cooling and internal chemical heating is applied). In a deterministic setup in only one time step the perfect action $\boldsymbol{u}_1$ could be decided to meet the target $\boldsymbol{S}_{\text{target}}$. We want to demonstrate, that this is more elaborate in a random variable framework.

In the context of this work, the application of random equation systems might be useful if there is only probabilistic knowledge about a) the initial state of the system $\boldsymbol{S}_1$, or b) the target state $\boldsymbol{S}_{\text{target}}$, or c) the parameters of the transfer functions $G$ and $H$. This is of special interest, if there are several distinct options for these variables (described by multi--modal mixture model densities) and still a singular decision about the action $\boldsymbol{u}_i$ must be taken at each time step. Then the first Equation (\ref{equ:control1}) is a random equation system with mixture models and we are interested in calculating the best action based on the posterior density of $\boldsymbol{u}_i$ where the prior of $\boldsymbol{u}_i$ may contain the information about the possible action space as described in Subsection \ref{sec:UsePrior}. For this purpose, we reformulate Equation (\ref{equ:control1}) to
\begin{align*}
\boldsymbol{G}(\boldsymbol{u}_i) + \boldsymbol{H}(\boldsymbol{S}_i) + \boldsymbol{S}_i  = \boldsymbol{S}_{\text{target}}\;.
\end{align*}
We reiterate the fact that with this formula, the control problem can have several target states by defining $\boldsymbol{S}_{\text{target}}$ as a mixture model and the control loop is working on to meet all those possible target states simultaneously. In order to meet the notation of the rest of the paper, we rewrite this to
\begin{align*}
G_r(\boldsymbol{x},\boldsymbol{A}_p) + H_r(\boldsymbol{A}_s,\boldsymbol{A}_p) + A_{s,r} = B_r\quad\forall\;r=1,..,R\;,
\end{align*} 
which is a random equation system with $K=R$ equations. Note, we subdivide the vector $\boldsymbol{A}$ in a) $\boldsymbol{A}_p$, which contains the random variables which are parameters of $G$ and $H$, and b) in the actual states $\boldsymbol{A}_s=\boldsymbol{S}_i$, that we want to alter by this control task.

\begin{figure}[htbp]
\centering
\includegraphics[width=9.5cm]{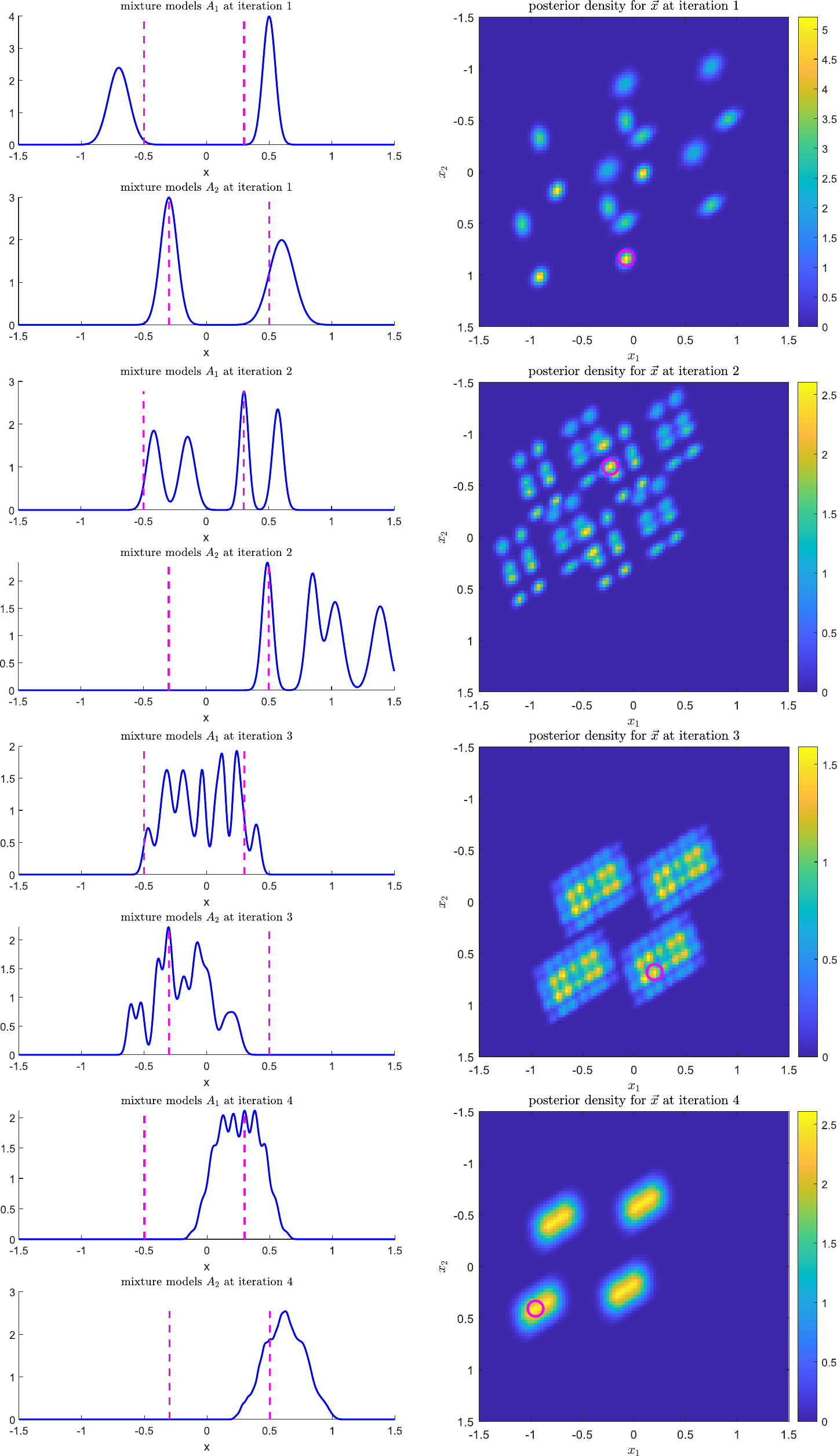}
\caption{Illustration of the control engineering problem for four iterations (iteration $1$ (top) to $4$ (bottom)). In each row, there is presented: The mixture model density of $A_1$ (top left) and $A_2$ (bottom left) as blue lines with additionally the two target states as dashed magenta line. Right: the posterior density of the random equation system showing the best action $\boldsymbol{x}$ for the next iteration as the maximum position with a magenta circle. The logic of the plots is as follows: Applying the maximum of the posterior in iteration $1$ leads to the mixture models $A_1$ and $A_2$ of iteration $2$ after applying the update. This is repeated until iteration $4$ where it can be observed that the $4$ target states are approximated simultaneously.}  
\label{fig:Res:control_001}
\end{figure}

In the following simulation study, we are utilizing a simple example for such a control engineering task. By defining $K=R=n=L_A=L_B=2$ (leading to four modes of the target state $\boldsymbol{S}_{\text{target}}$) and the deterministic linear functions (i.e. there is no $\boldsymbol{A}_p$)
\begin{align*}
\boldsymbol{G}(\boldsymbol{x}) := \left(\begin{array}{cc} 1 & \gamma \\ \gamma & 1  \end{array}\right)\cdot \boldsymbol{x}\;,\quad \boldsymbol{H}(\boldsymbol{A}_s) := \left(\begin{array}{cc} -\alpha & \beta \\ -\beta & -\alpha  \end{array}\right)\cdot \boldsymbol{A}_s
\end{align*}
we get the random equation system
\begin{align*}
x_1 + \gamma\,x_2 + (1-\alpha)\,A_{s,1} + \beta\,A_{s,2} &= B_{r,1}\\
\gamma\,x_1 + x_2 - \beta\,A_{s,1} + (1-\alpha)\,A_{s,2} &= B_{r,2}\;,
\end{align*}
for finding the best action $\boldsymbol{x}$ and the corresponding update equations
\begin{align}
A_{s,1}^{\text{next}} &:= x_1 + \gamma\,x_2 + (1-\alpha)\,A_{s,1} + \beta\,A_{s,2} \label{equ:control_sim_up1}\\
A_{s,2}^{\text{next}} &:= \gamma\,x_1 + x_2 - \beta\,A_{s,1} + (1-\alpha)\,A_{s,2}\label{equ:control_sim_up2}\;.
\end{align}
If we start in the first iteration with a Gaussian mixture model for $A_{s,1}$ and $A_{s,2}$ with $L_A=2$ components, then we get in the second iteration (by the application of the first update with a decided action in Equations (\ref{equ:control_sim_up1}) and (\ref{equ:control_sim_up2})) at $L_A^2$ Gaussian components. Along this line of argumentation, the number of mixture model components grows quadratically from iteration to iteration. On the other hand, by selecting action $\boldsymbol{x}$ as the maximum of this multi--modal posterior, we work against the internal dynamic of the system $\boldsymbol{H}(\boldsymbol{A})$ in order to approximate the (possibly multi--modal) target state $\boldsymbol{S}_{\text{target}}$ as much as possible. As stated, in this example we have $L_B^R=4$ different mono--modal combinations for the target state. In Figure \ref{fig:Res:control_001} a simulation example is presented with $\gamma=0.2,\alpha=0.4,\beta=0.3$ for four iterations (top to bottom). In the top row we start with $L_A=2$ Gaussian mixture model components which leads to {$16$} peaks in the posterior for the solution of the equation system in the first iteration. This number comes from the fact that one has $L_A^K\cdot L_B=8$ options to select $x_1 + \gamma\,x_2$ in the first equation to achieve a high likelihood and the same amount for $\gamma\,x_1 + x_2$ in the second equation, leading to {$2\cdot 8=16$} combinations for $(x_1,x_2)$. In the last plot (fourth iteration) we already have $16^2=256$ Gaussian components in each mixture model density of $A_{s,1}$ and $A_{s,2}$. Due to the squaring of Gaussian components from iteration to iteration the densities have the potential to be more volatile (note, at iteration $5$ we already have $65536$ mixture model components in $A_{s,1}$ and $A_{s,2}$), but due to $\alpha,\beta\in[0,1]$ we see a contraction of the overall standard deviation by a strong overlapping of Gaussian components, approximating the four target states in $\boldsymbol{S}_{\text{target}}$ with a convergence to four possible actions as seen in the posterior plot on the right. Of course, there is no need to calculate this quadratically growing number of mixture model components explicitly, since a) one can simply work with an overall approximation to the state densities in this loop and b) although the number of components is growing fast, they are converging which makes the large number of parameter combinations less distinguishable as presented in Figure \ref{fig:Res:control_001} (bottom row).

\textit{Observation \#{}8:} Random equation systems with mixture models can be beneficially utilized in iterative schemes and optimizing for different targets simultaneously, for example in control problems. This shows the potential to model time--discrete dynamic processes with several simultaneous states.

\subsection{Application to Random Matrix Theory}
\label{sec:res:matrix}

Random matrix theory proved to be a very useful approach investigating physical properties of quantum systems \cite{Livan2018} or performing statistical analysis \cite{Debashis2014}. A very important task in this theory is to calculate the eigenvalues of a random matrix \cite{Bharucha-Reid1972}, i.e. a matrix that contains random variables, such as for a real and symmetric $3\times 3$ matrix
\begin{align*}
M := \left(\begin{array}{ccc} A_{11} & A_{12} & A_{13} \\ A_{12} & A_{22} & A_{23} \\ A_{13} & A_{23} & A_{33}\end{array} \right)
\end{align*}
which leads to the characteristic polynomial $p(x):=\text{det}(M-x\,I)$ of the eigenvalue problem
\begin{align*}
p(x)=\;& -x^3 + (A_{11} + A_{22} + A_{33})\cdot x^2 \\
&- (A_{11}A_{22} + A_{11}A_{33} + A_{22}A_{33} - A_{12}^2 - A_{13}^2 - A_{23}^2)\cdot x\nonumber\\
& + (A_{11}A_{22}A_{33} + 2A_{12}A_{23}A_{13} - A_{13}^2A_{22} - A_{11}A_{23}^2 - A_{12}^2A_{33})\;.
\end{align*}
In order to find the eigenvalues, the roots of this polynomial have to be determined, i.e. $p(x)=0$ (again, we will utilize the argumentation framework as described in Subsection \ref{sec:implicit}). In general, for a real symmetric $n\times n$ matrix with $L_A$ mixture model components in each matrix entry there are $L_A^{\frac{n(n+1)}{2}}$ different combinations of mono--modal random matrices. This already leads for $L_A=6$ and $n=15$ to  $6^{120}\approx 2.4\cdot 10^{93}$ different mono--modal random matrix configurations (the same number as for the last example of intersections of conic sections in Subsection \ref{sec:res:randomconic}).

\begin{figure}[htbp]
\centering
\includegraphics[width=11cm]{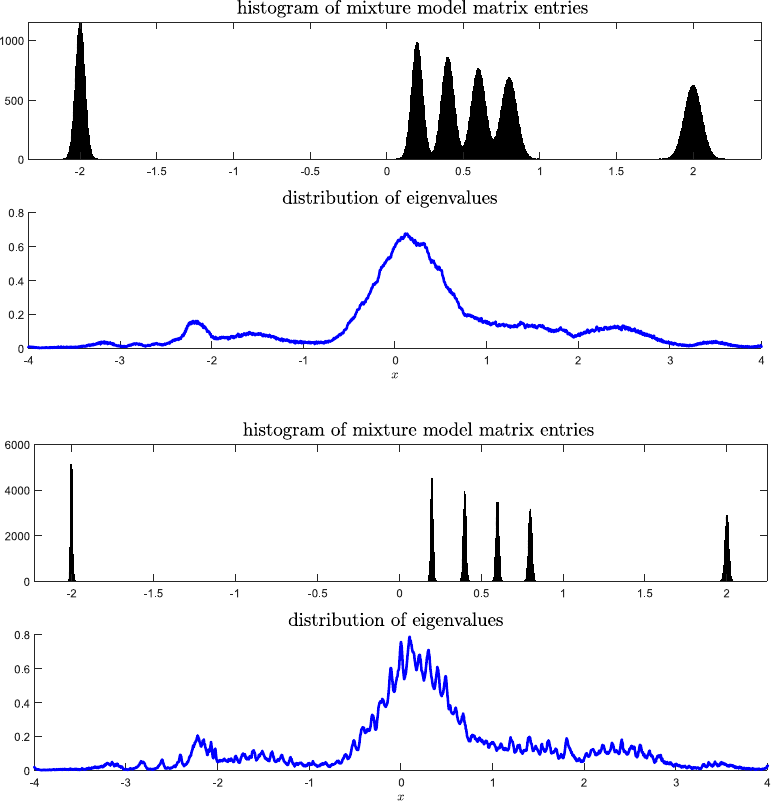}
\caption{Illustration of the combinatorial eigenvalue problem for two mixture model component standard deviation ranges ($\sigma\in[0.3,0.6]$ (top) and $\sigma\in[0.006,0.012]$ (bottom)). In each row, there is presented: Top: The histogram of the samples of all the identical matrix entry mixture model random variables. Bottom: The posterior density of the eigenvalues (utilizing a constant prior).}  
\label{fig:Res:eigvalues_001}
\end{figure}

For numerical simplicity we want to present simulation results for the $3\times 3$ case and utilize $L_A=6$ leading to $L_A^{3\cdot 4/2} = 6^6 = 46656$ different mono--modal random matrices with their corresponding eigenvalue densities. One typical question in random matrix theory could be what the spectral density is in the case for a specific set of mixture model components for $A_{ij}$ leading to this combinatorial setup of random matrices. For example, this can help to determine the definiteness of a specific combinatorial matrix family. Such questions can be approximated in a direct fashion with the concepts presented, and in Figure \ref{fig:Res:eigvalues_001} we utilize for all entries the same Gaussian mixture models for the $3\times 3$ case (with arbitrarily chosen mode positions at $\{-2,0.2,0.4,0.6,0.8,2\}$) but different component standard deviations $\sigma\in[0.3,0.6]$ (top) and $\sigma\in[0.006,0.012]$ (bottom). Although in the top row we already see concentrations and peaks of the eigenvalue density due to the multi--modal structure of the random variables, a much more detailed structure gets visible by reducing the individual component widths indicating a more discrete nature of the spectral density. In this example, we only want to demonstrate that such calculations can be done efficiently with the framework presented and are not focusing on specific applications of mixture models in random matrix theory. An interesting question that might arise in random matrix theory in this context could be: How can \textit{Wigner's semicircle law} be approximated by such a composition of mixture model random variables in a random matrix?

\textit{Observation \#{}9:} The utilization of this argumentation and efficient computation framework may lead to new applications and research questions in neighboring research fields, such as in random matrix theory.

\section{Discussion}

We demonstrated the combinatorial potential of nonlinear random equation systems with mixture models in a Bayesian argumentation framework and presented simplified examples how such equations can be utilized in different areas of application. We see this approach in the tradition of dealing with combinatorial complexity by compact formulations such as, for example, utilized for the \textit{kernel trick} of machine learning with high dimensional feature spaces \cite{Schoelkopf1998}. The perspective of this presentation is to allow a broad applicability by general derivations (no significant constraints about the type of equations and probability densities) as well as a straightforward argumentation, numerical considerations and expressive simulation experiments for the applied researcher. We consider this practical perspective to be the main character of this presentation, which should inspire further applications in the future.

Since we took a straightforward modeling approach, questions of more complicated dependency structures of random variables are not specifically addressed. Of course, the derivation of the likelihood and Monte Carlo integration in Equation (\ref{equ:MonteCarloInt}) can also be performed for stochastically dependent entries of the parameter vector $\boldsymbol{A}$, although this could strongly decrease the numerical efficiency. In this context, working with random equations can be regarded as part of the very broad field of uncertainty quantification which gained a lot of attention in the recent years \cite{Sullivan2015}. The specific perspective of uncertainty quantification could be: Finding best solutions of (nonlinear) equation systems, which parameters contain uncertainties understood as limited knowledge about their values, and investigating how these solutions depend on different parametrizations of these uncertainties.

The type of mixture models utilized in this presentation is unusual, since in most applications mixture models do not occur (or even are constructed) with disjoint components. Although we do not assume disjoint mixture model components in this work, we certainly have at least well separated modes of such densities in mind when discussing the combinatorial complexity. Certainly, the presented derivations are also valid for highly overlapping mixture model components, but for such a case different applications would be the focus.

The presented examples have the goal to show in simple, well--arranged simulations several important characteristics for this type of modeling which are summed up by \textit{observations \#{}1--9}. These observations show that even minor changes in modeling of REs with mixture models may lead to major changes in the solution space and capture how this leads to unexpected effects, specifically in example applications containing optimization, iteration schemes or random matrix analysis. 

In this work, we demonstrate and illustrate the likelihood function and posterior density in simulation studies and are not focusing on the calculation of point estimates based on the likelihood or posterior, such as Maximum Likelihood (ML), Maximum A Posteriori (MAP) or the expectation value of the posterior by Minimum Mean Squared Error (MMSE) \cite{Hoegele_2013}. As we demonstrate in the results section the calculation of these best estimates for $\boldsymbol{x}$ could be difficult, since the extrema of the resulting function are not necessarily distinct or unique (challenging for ML and MAP) and the domain of non--zero posterior density could be large while containing small scale structures (challenging for MMSE). This results directly from the application of mixture models and is inherent to the investigated random equations. Even further, an attractive approach can be to work with the full probabilistic solution space represented by the posterior density and utilize it directly in further investigations, without calculating point estimates. 

{We are optimistic that this broad presentation will prove practically useful across many fields, as demonstrated by the introductory application examples in Subsections \ref{sec:res:portfolio}, \ref{sec:res:control}, and \ref{sec:res:matrix}. From our perspective, any application involving an intractable combinatorial component and seeking approximate solutions to equations or optimization problems should generally be adaptable to the framework we have presented. Consequently, we anticipate numerous future applications building upon this work.}

Although we are not experienced in quantum computing, we see a certain potential of the use of mixture model random variables in order to calculate superpositions of states in a simple way. Similar to quantum computing, we see a major opportunity to calculate expressions with inherently high combinatorial components. In this respect, the similarities and differences of quantum linear equation systems \cite{dervovic_2018} compared to the random linear equation systems with mixture models may help to establish a connection between these two fields.

In total, the presented derivations and simulations should be regarded as an introductory presentation for this new type of equations based on REs with mixture models and future work for specific application areas is encouraged.

\section{Conclusion}

We have demonstrated how nonlinear equation systems with uncertainties and high combinatorial complexity in the parameters can be modeled and approximately solved by simulations. This task is introduced by solving random equations with mixture models with (practically) disjoint components. Approximate solutions can be efficiently calculated by a general derivation of the likelihood function with no significant constraints about the type of equation and probability densities. Utilizing a Bayesian framework, we calculate the posterior of the solution, which might lead to new insights in many different practical applications. We have presented numerical example results for random linear equation systems, systems of random conic section equations, as well as applications to portfolio optimization, stochastic control and random matrix theory. These example results show the wide applicability of the methodology presented in this study as well as its efficient computation.

\section*{Acknowldegments}
Many thanks for the discussions and advices to Associate Professor Dr. Michael A. Hoegele at the \textit{Department of Mathematics, Los Andes University, Bogotá, Colombia}.

% Bibliography

\end{document}